\documentclass[fleqn,10pt]{wlscirep}
\usepackage[utf8]{inputenc}
\usepackage[T1]{fontenc}
\usepackage{amsmath}

\title{Modelling lifespan reduction in an exogenous damage model of generic disease} 

\author[1,2]{Rebecca Tobin}
\author[1]{Glen Pridham}
\author[1,*]{Andrew D. Rutenberg}
\affil[1]{Department of Physics and Atmospheric Science, Dalhousie University, Halifax, Nova Scotia, Canada, B3H 4R2}
\affil[2]{Data Science, Analytics, and Artificial Intelligence (DSAAI) program, Carlton University, Ottawa, Canada, K1S 5B6}

\affil[*]{adr@dal.ca}

\begin{abstract}
We model the effects of disease and other exogenous damage during human aging. Even when the exogenous damage is repaired at the end of acute  disease, propagated secondary damage remains. We consider both short-term mortality effects due to (acute) exogenous damage and long-term mortality effects due to propagated damage within the context of a generic network model (GNM) of individual aging that simulates a U.S. population. Across a wide range of disease durations and severities we find that while excess short-term mortality is highest for the oldest individuals, the long-term years of life lost are highest for the youngest individuals. These appear to be universal effects of human disease. We support this conclusion with a phenomenological model coupling damage and mortality. Our results are consistent with previous lifetime mortality studies of atom bomb survivors and post-recovery health studies of COVID-19. We suggest that short-term health impact studies could complement lifetime mortality studies to better characterize the lifetime impacts of disease on both individuals and populations.
\end{abstract}

\begin{document}
\flushbottom
\maketitle

The emergence of novel diseases -- such as COVID-19, Ebola, SARS, Zika, avian flu, or monkeypox -- is a worsening trend.\cite{Jones:2008} Every new disease raises urgent questions about how they could impact infected individuals and the population at large. Yet observational studies offer answers only in retrospect. How can \emph{a priori} knowledge inform us \emph{before} new diseases are studied and characterized? One approach is to identify potentially universal effects of disease. This approach may also be useful for existing diseases that are not yet fully characterized.

Rapidly increasing mortality with age of infected individuals is a common feature of many infectious diseases.\cite{Lu2020-fg, Wong:2004, Thompson:2003, Simonsen1998-gs, Agua-Agum2015-qg, Boelle:2002, Wang:2019, Gil:2004} For example, short-term mortality due to COVID-19 rises approximately exponentially  with age -- more than 30-fold from 55 to 85 years.\cite{Levin:2020, Goldstein:2020}  Many infectious diseases also exhibit long-term complications, exemplified by post-acute `sequelae' (PAS) -- for example, SARS and MERS,\cite{Ahmed:2020} Ebola,\cite{Wilson:2018} Zika,\cite{Souza:2019} `long COVID',\cite{Thompson:2022} and COVID complications.\cite{Mulberry2021-mq} Surprisingly, we do not know the long-term effects of most PAS,  how they depend on age, or how they compare to the impact of short-term mortality. This is because there are very few long-term, large-scale studies of the impact of acute disease; most studies are limited to less than 5 years. One notable exception is the study of lifetime mortality impacts of exposure to the atomic bombs at Hiroshima and Nagasaki.\cite{Preston:2003, Ozasa:2012} While this does not represent the effects of disease, it does represent the long-term effects of acute exogenous damage. 

Understanding age-effects of disease is particularly important. For example, assuming that short-term mortality is the only impact of acute diseases implies that immunization of older individuals will typically \cite{Bubar2021-kw} save more years of life than immunizing younger individuals.\cite{Goldstein:2020, Goldstein:2021} However, if post-acute health impacts of disease -- including PAS -- lead to substantial shortened lifespans then immunizing \emph{young} individuals could save more years of life. Resolving these questions of age-effects for individual diseases is not easily done, since lifetime observational studies require many decades. 

A promising \textit{a priori} approach is to computationally model the age-effects of disease. This first requires a model of normal aging. Encouragingly, aging populations exhibit simple and universal behavior. Average human mortality rates exhibit an exponential increase with age known as Gompertz' law,\cite{Kirkwood:2015} which is reminiscent of the increased short-term mortality of disease with age. Individual health can be captured by the frailty index, which measures damage and dysfunction.\cite{Searle:2008}
Before death, individuals accumulate damage approximately exponentially with age,\cite{Mitnitski:2001} leading to worsening individual health.\cite{Howlett:2021} 
The random but inexorable accumulation of damage during aging can be modelled at the individual level by a complex network of binary health attributes (healthy or not),\cite{Gleeson:2013} where damage propagates stochastically across static links (edges).  Such a ``Generic Network model'' (GNM) of human aging recovers the population-level behaviour of mortality and health.\cite{Farrell:2016, Mitnitski:2017, Farrell:2018, Rutenberg:2018} 

A GNM model provides a dynamical context for propagating damage due to disease. We can model the onset of disease by treating it as an exogenous event that further damages an individual. As such, we can also consider any exogenous damage -- and are not specifically limited to  disease. While the generic nature of the health attributes in the GNM precludes a detailed study of specific diseases, its generic nature allows us to identify and characterize potentially universal effects of disease in aging individuals. 

We will consider the effects of disease timing (onset age), severity, and duration.  We will first consider excess mortality (fatality) rates due to disease. To assess the long-term impact of diseases we also need to consider years of life lost due to damage originating from disease. We can use years of life lost within different time horizons to compare short and long-term impacts of disease. We also develop and explore a simplified phenomenological model of how exogenous damage leads to earlier mortality.

\section*{Generic Network Model (GNM) of  Disease and  Exogenous Damage}
The GNM represents individual health by an undirected scale-free network.\cite{Stubbings:2023} Links, defining network topology, are static. Nodes are dynamic binary health attributes -- either damaged or not. A summary measure of individual health is the frailty index ($f$),\cite{Searle:2008, Howlett:2021} which is the fraction of damaged nodes.  An undirected scale-free network is generated using the  Barab\'{a}si-Albert preferential attachment model,\cite{Barabasi_Albert:1999} with an average node degree $\langle k \rangle$ and scale-free exponent $\alpha_{GNM}$. Nodes are initially undamaged at age $t=0$, but damage at a rate  $\Gamma_+ = \Gamma_0 \exp(\gamma_+ f_i)$, where $f_i$ is the fraction of damaged neighbours for node $i$. Damaged nodes repair at a rate $\Gamma_- = (\Gamma_0/R) \exp(\gamma_- f_i)$, though repair has a negligible effect on population statistics in practice. Individual mortality occurs when the two most connected nodes are both damaged. We use previously determined GNM parameters\cite{Farrell:2016, Farrell:2018} that approximate sex-combined USA population health and mortality statistics\cite{Arias:2014} for ages $t \gtrsim 20$: $\langle k \rangle=4$, $\alpha_{GNM}=2.27$, $\Gamma_0$ = 0.00183, $\gamma_+$ = 7.5, with small repair ($\gamma_- = 6.5$ and $R = 3.0$) and $N=10^4$ nodes. For simplicity and clarity we do not use a false-negative correction \cite{Farrell:2016} to reduce the range of $f$ to $[0,1-q]$  -- i.e. we use $q=0$ and have $f \in [0, 1]$. Stochastic dynamics are exactly sampled.\cite{Gillespie:1977} All plotted data corresponds to at least $10^6$ simulated individuals. Errorbars for averages, unless indicated, are smaller than point sizes. All times are in years. 

The GNM models damage from all sources that arises during the aging process, including the propagation or amplification of earlier damage. It then captures mortality effects due to that damage. Since the GNM is parameterized from population health and mortality statistics, it implicitly includes many extrinsic events such as disease or injury -- the usual stressors of living. As such we expect that the GNM will allow us to model the effects of an individual disease, which we here consider as additional or perturbative to the normal aging process in order to estimate its effect. 

\begin{figure}[t] 
    \includegraphics[width=.9\textwidth]{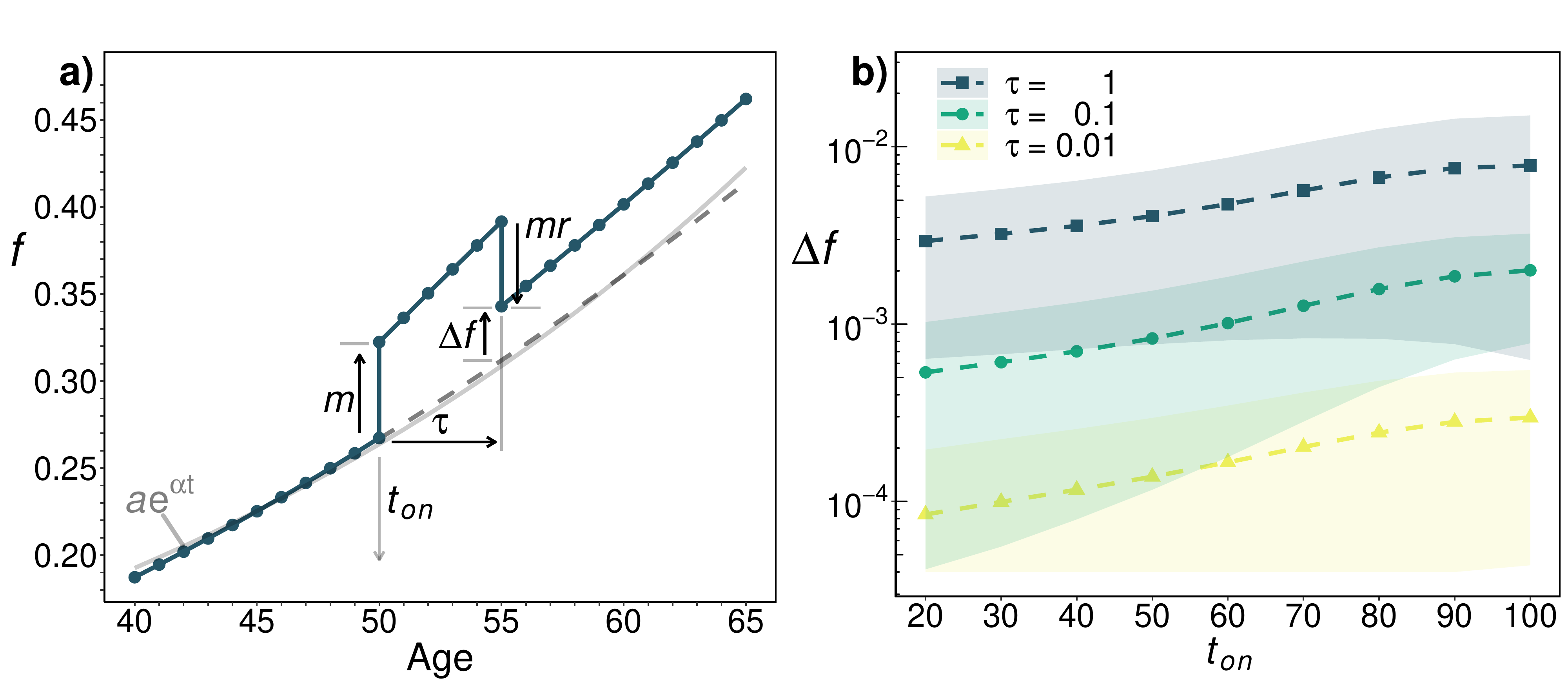}
    \caption{ {\bf(a)  Model disease.} A disease is represented by exogenous damage of severity $m$ inserted at onset time $t_{on}$; a fraction $r$ of the original damage is then removed after duration $\tau$. Excess damage that is left at $t_{on}+\tau$ is indicated by $\Delta f$. The average damage vs age, as assessed by the frailty index ($f$, the fraction of damaged nodes within the GNM), for an acute disease with $r=1$, $m=0.05$, $t_{on}=50$ and $\tau=5$ is indicated by the blue points. A control population (with $m=0$) is indicated by the grey dashed line, and is well approximated by an exponential $f=a e^{\alpha t}$ where $a=0.0548\pm0.0009$, $\alpha = 0.0314\pm0.0003$, and $t$ is the age -- as indicated by the solid grey curve. 
      {\bf (b) Excess damage.} Increase in the frailty index at the end of an acute disease, $\Delta f$ at $t=t_{on}+\tau$, with severity $m = 0.02$ vs onset age $t_{on}$, with duration $\tau$ as indicated by legend and $r=1$. The shading indicates the standard deviation of $\Delta f$. All ages and times, in this and other figures, are in years.
}
    \label{fig1}
\end{figure}

We will not model details of the disease process, rather we will simply assume the disease starts (e.g. due to infection) at some onset age $t_{on}$ and lasts for a duration $\tau$.  In a similar spirit we will assume that the disease has a fixed severity or magnitude $m$. In terms of the GNM, our model disease damages a fraction $m$ of nodes at the onset age $t_{on}$.  While formally $m \in [0,1]$, we do not damage already damaged nodes so $m$ is kept small. We exclude individuals from analysis who have initial damage $f > 1-m$. For $m \leq 0.02$ no individuals are excluded, while for $m = 0.05$ a small fraction ($10^{-4}$) are excluded for $t_{on} \geq 90$. At the end of the disease (at $t_{on}+\tau$) a fraction $r$ of the applied damage is removed. The fraction $r$ of damage that is removed is a recovery or ``resilience'' parameter. For acute diseases we typically use $r=1$, while chronic diseases could be modelled with $r=0$ (equivalently, $\tau \rightarrow\infty$). Since we model disease by introducing exogenous damage $m$ at time $t_{on}$, and allow for a fraction $r$ to be repaired after $\tau$ through resilience, we can use the same model for any exogenous damage. The effect of our model disease is illustrated in Fig.~\ref{fig1}a with respect to the frailty index $f$. The control population with no disease is indicated by the grey dashed line. We see that even with $r=1$ there is excess damage $\Delta f$ left in the individual after the end of the disease. This residual damage leads to long-term mortality effects that we characterize.  We compare these long-term effects with the short-term acute effects that we also characterize. 

We measure long-term mortality using the average reduction in lifespan ($\Delta t_{tot}$) and also by the average years lost within a window of $w$ years after the disease ($\Delta t_w$), assuming the mortality rate of the control population after that window. All disease results are with respect to a large control population with no disease ($m=0$). The excess probability of death due to the disease corresponds to an excess Infection Fatality Rate (IFR) as compared to the control population. 

\begin{figure}[t]  
   \includegraphics[width=.9\textwidth]{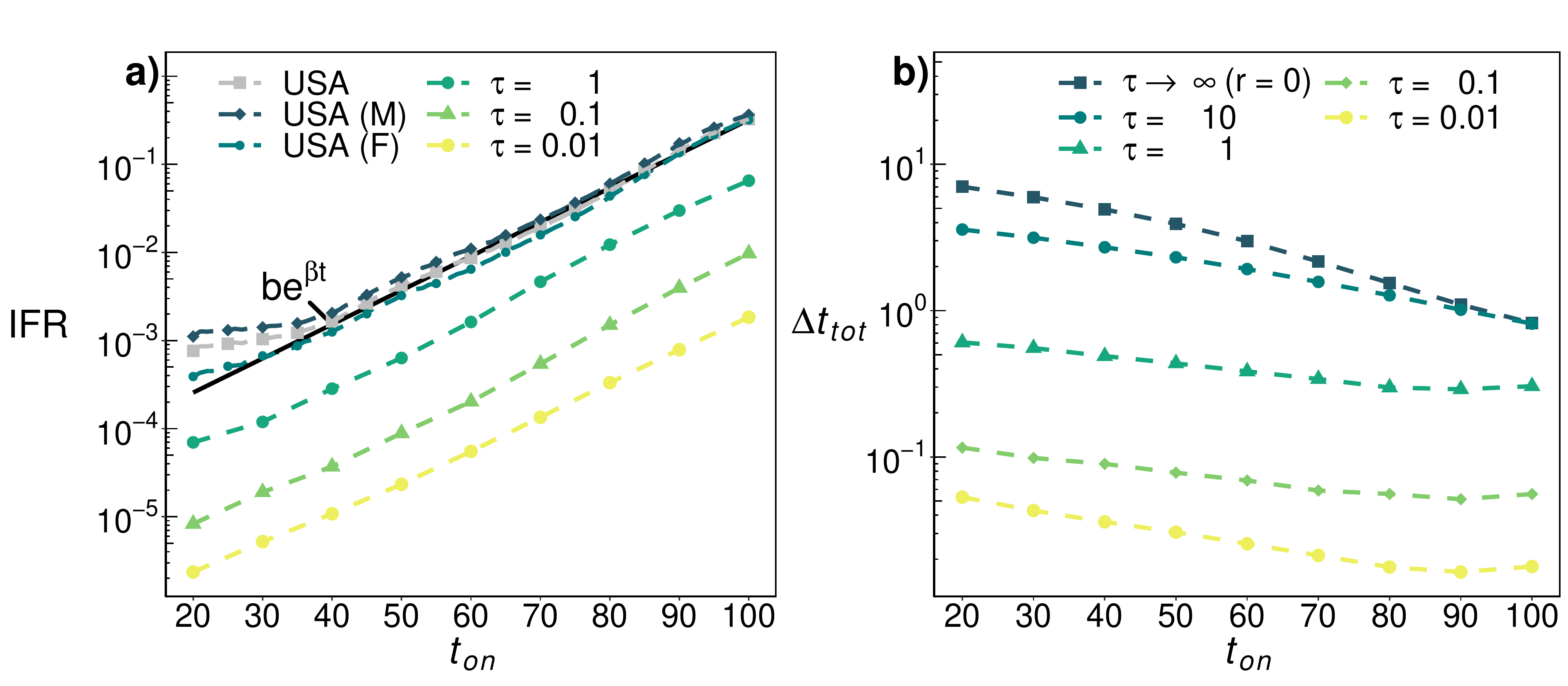}
   \caption{{\bf (a) Mortality}.  Excess probability of death during the disease (IFR) vs onset age ($t_{on}$) for acute diseases with duration $\tau$ as indicated, and $m=0.02$. Square grey markers indicates the all-causes mortality rate (per year) vs. age from the U.S. population (2010).\cite{hmd} Exponential fit (solid black line): $(4.3 \pm 0.3) \times 10^{-5}\exp{[(0.089\pm0.001)t_{on}]}$.
   Male (M) and female (F) sub-populations are as indicated. 
     {\bf (b) Lifespan reduction}. The average total reduction in lifespan due to disease, $\Delta t_{tot}$, vs. onset age $t_{on}$ for severity $m = 0.02$ and duration $\tau$ as indicated by legend, with $r=1$. Chronic disease corresponds to $\tau=\infty$ (or $r=0$).
}
    \label{fig2}
\end{figure}

\section*{GNM Results}
Our GNM model disease has a significant impact on long-term health, as shown by the average frailty index ($f$) vs age for large simulated populations that received a disease (blue points and solid line) or did not (grey dashed line) in Fig.~\ref{fig1}a.  With maximal resilience ($r=1$, our default acute disease) all of the damage introduced at $t_{on}$ is removed after $\tau$. Nevertheless excess damage propagates within the GNM and remains at $t_{on}+\tau$, as indicated by $\Delta f$. For a variety of onset ages, and for selected durations $\tau$ as indicated, we show $\Delta f$ in Fig.~\ref{fig1}b. We see that $\Delta f$ increases with onset age, and also that the individual variability of propagated damage (indicated by the shaded regions) is  large. This reflects the stochastic nature of damage propagation within the GNM.

In Fig.~\ref{fig2}a, we show the excess mortality during an acute disease (IFR) vs onset age $t_{on}$. The IFR increases monotonically with $t_{on}$ for all $m$ and $\tau$ investigated, and maintains an approximately exponential age dependence similar to the all-causes mortality curve ($\mu$, grey squares). In Fig.~\ref{fig2}b we show the total years lost due to disease ($\Delta t_{tot}$) vs the onset age. Strikingly, we see that the average reduction in lifespan is highest for younger populations (note the log-scale). There are two mechanisms that could contribute to the reduction of lifespan of younger individuals. The first is that mortality during the disease leads to more years of life lost for younger individuals -- who have more years left in their life expectancy. The second is that long-term mortality effects could be worse for younger individuals. We can separate these effects by considering  different observation windows $w$ after the disease.

\begin{figure}[t] 
    \includegraphics[width=.9\textwidth]{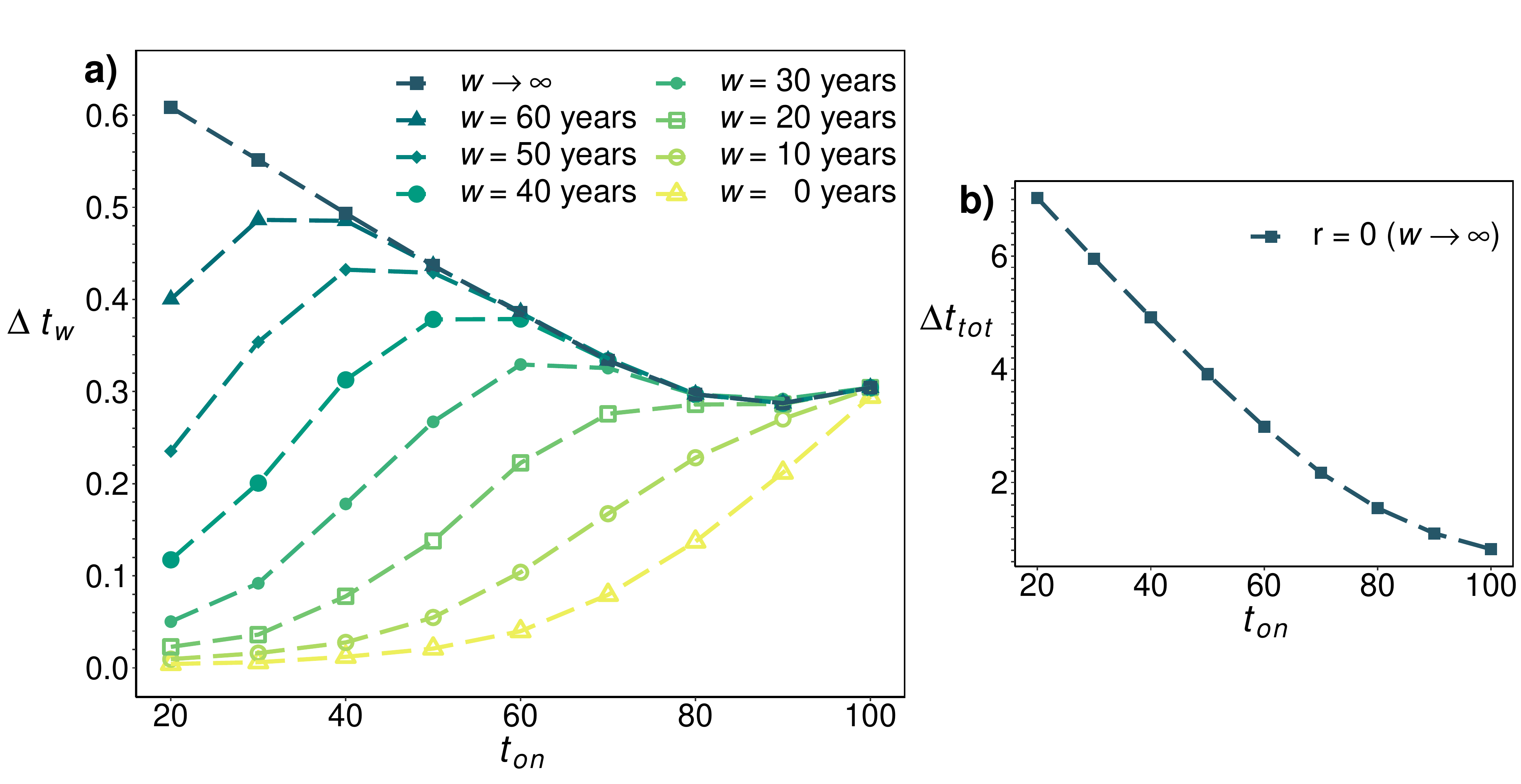}
    \caption{ {\bf Lifespan reduction for different observation windows. } \textbf{(a)} The average years lost $\Delta t_w$ vs $t_{on}$ for different observation windows $w$ past the end of acute disease (with $r=1$). The effects of mortality during the disease ($w=0$) are largest for older individuals, even though the younger individuals have more lifespan left to lose. The effects of lifetime mortality ($w \rightarrow \infty$) are largest for younger individuals, demonstrating the impact of residual damage. All with $\tau$ = 1 and $m$ = 0.02.  \textbf{(b)} $\Delta t_{tot}$ for a chronic disease ($r=0$). The lifetime effects ($w \rightarrow \infty$) are much larger than in Fig.~\protect\ref{fig3}a.
    }
    \label{fig3}
\end{figure}

In Fig.~\ref{fig3}a we show the average years lost $\Delta t_w$ within a window of duration $w$ after the end of the disease. We account for all excess mortality between $t_{on}$ and $t_{on}+\tau+w$. Just considering deaths during the disease ($w=0$, yellow open triangles), we find that older populations have the largest number of years lost -- as observed with, e.g.,  COVID-19.\cite{Goldstein:2021} Even though younger individuals have more lifespan left to lose, it is not enough to offset their much lower IFR. However, for younger ages years lost due to deaths during the disease account for only a small fraction of the total years lost. As we increase $w$, $\Delta t_w$ increases, and its peak shifts towards younger ages.  The largest lifetime impact ($\Delta t_\infty \equiv \Delta t_{tot}$, blue squares) is for the youngest individuals, in agreement with Fig.~\ref{fig2}b. This effect holds for a wide range of $\tau$ and $m$ parameter values, see Supplemental Figs.~S2 and S3. Strikingly, the peak (mode) of lifespan impact only moves away from the oldest ages with long observation windows of $w \gtrsim 20$ years. The ratio of lifespan reduction $\Delta t_{tot}/\Delta t_0$ exceeds $100$ for the youngest onset ages, and does not strongly depend on duration $\tau$ or severity $m$ (Supplemental Fig.~S1). The ratio will further increase for lower resilience ($r<1$) since acute mortality, IFR, and acute life lost, $\Delta t_0$, are unchanged but mortality after the disease is increased due to larger residual damage $\Delta f$. For example, in Fig.~\ref{fig3}b with $r=0$ we show that $\Delta t_{tot}$ is more than ten-fold larger than with $r=1$.

\section*{Phenomenological Model of Disease and Exogenous Damage} 
While the GNM allows for stochastic and high-dimensional individual health trajectories, the connection between modelling assumptions and phenomenological behavior is obscured by its complexity. A simpler model would be more interpretable -- allowing us to see how and when our modelling assumptions lead to the behavior we see. A simpler model would also be easier to generalize. While other mean-field versions of the GNM exist,\cite{Farrell:2016, Farrell:2018} here we develop a simple model that is directly rooted in the observed aging phenomenology: damage accumulates non-linearly with age and this damage drives mortality. The essential simplification here is that the health-state is described only by the average damage -- rather than by the many interconnected nodes of the GNM. This phenomenological model complements our network-based simulations using the GNM, and can be easily modified for different phenomenological assumptions. 

We start with the observation that the average damage, or frailty index, increases approximately exponentially with age $f_0(t)=a e^{\alpha t}$. \cite{Mitnitski:2015} From the GNM, we have $\alpha \approx 0.031$ (and $a \approx 0.055$, see Fig.~\ref{fig1}) which is consistent with observational estimates for adults with $t \gtrsim 20$ ($\alpha \approx 0.035 \pm 0.02$ \cite{Mitnitski:2015}). We assume that exogenous damage, such as from  disease or injury, forms part of --- and behaves similarly to --- the damage exhibited during aging. As such it satisfies the differential equation $df/dt = \alpha f$ and any exogenous damage $m$ grows exponentially thereafter. By including resilience, we then have simple expressions for the average damage before, during, and after the disease:
\begin{align}
f(t) &=
\begin{cases}
a e^{\alpha t} \, \, &~~~~~~~~~t < t_{on},\\
a e^{\alpha t}+m e^{\alpha (t-t_{on})} \, \, &t_{on} < t < t_{on}+\tau, \\
a e^{\alpha t} + \Delta f \, e^{\alpha (t-(t_{on}+\tau))} \, \, &~~~~~~~~~t> t_{on}+\tau,
\end{cases}
\label{eq:f(t)}
\end{align}
where 
\begin{align}
\Delta f = m (e^{\alpha \tau}-r)
\label{eq:deltaf}
\end{align}
is the propagated damage at the end of the acute disease (at $t_{end}=t_{on}+\tau$, and with resilience $r$).  

This phenomenological damage model is already considerably simplified compared to the GNM: we have a single deterministic health state variable ($f$) rather than $N=10^{4}$ distinct and stochastic health-nodes. By comparing our expression for the propagated damage $\Delta f$ (Eqn.~\ref{eq:deltaf}) with Fig.~\ref{fig1}b, we see that the phenomenological model has a single value of $\Delta f$ that is independent of onset age $t_{on}$ while the GNM has a broad range of $\Delta f$ with an average that increases with $t_{on}$ -- though by much less than the individual variability. 

We also need an explicit mortality model. We use the well-established but phenomenological Gompertz law of $\mu_0 = b e^{\beta t}$,\cite{Gavrilova:2015} whereby the mortality rate of adults increases exponentially with age. We estimate $\beta \approx 0.089$ (and $b\approx 4.3 \times 10^{-5}$, see Fig.~\ref{fig2}a). We then assume that the increasing mortality rate results \emph{only} from the increasing frailty-index $f(t)$. To obtain the correct time-dependence for mortality from $f_0 \propto e^{\alpha t}$ we  have 
\begin{align}
\mu = b (f/a)^{\beta/\alpha}.
\end{align}
This expression will hold for both the disease and control populations, since by assumption the mortality is expressed only through the health. With a disease, for $t>t_{end}$ we can express this as 
\begin{align}
&& \mu(t) = b e^{\beta t} \bigg(1+\frac{\Delta f}{f_{end}}\bigg)^{\beta/\alpha},
\label{eq:hazard}
\end{align}
where $f_{end} = f_0(t_{end})=a e^{\alpha (t_{on}+\tau)}$ is the control (non-disease) frailty at the end of the disease. Note that a chronic disease corresponds to a disease with no resilience -- i.e. $r=0$. A similar expression for the hazard applies during the disease, with the ratio $\Delta f/f_{end}$ replaced by $m/f_{on}$. 

\begin{figure}[t] 
    \centering
\includegraphics[width=.9\textwidth]{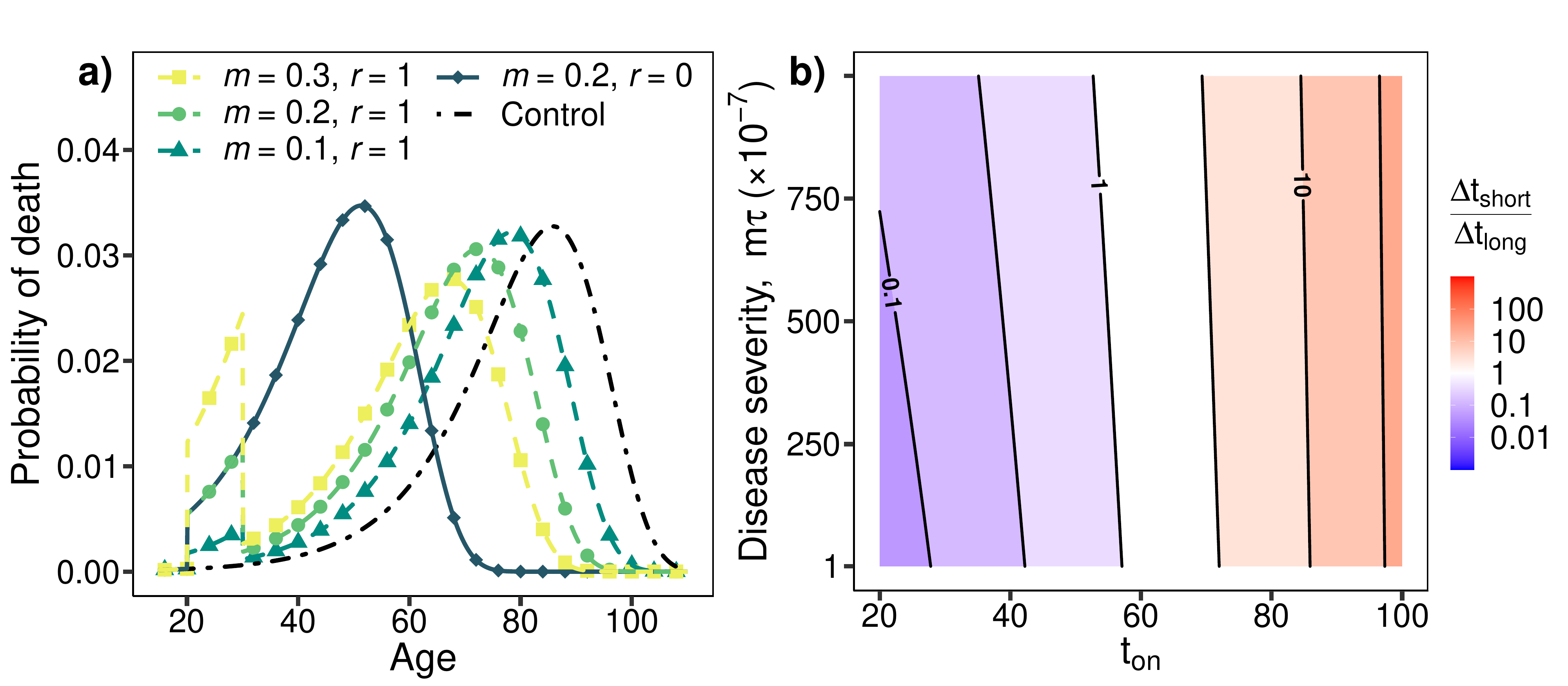}
    \caption{ {\bf Phenomenological model. } \textbf{(a) Effect of varying $m$ and $r$ on death age}. The control distribution (black, dot-dashed line) is shifted towards lower ages by the disease.  With resilience (dashed lines), two phases emerge: an acute phase during the disease (ages 20-30) and a chronic phase after the disease ends, due to propagated damage. Each phase contributes to the overall loss of life due to the disease. Without resilience (solid line, $r=0$) the two phases merge into a single short-lived persistent phase. ($\tau = 10$, $t_{on}=20$) \textbf{(b) Acute vs chronic effects}. Ratio of expected life lost during acute phase vs chronic phase, $\Delta t_{short}/\Delta t_{long}$. The ratio increases approximately exponentially with increasing age of onset, $t_{on}$, nearly independently of disease severity ($m\tau$). ($\tau = 10^{-3}$, $10^{-4} \leq m \leq 10^{-1}$, $r=1$) 
    }
    \label{fig4}
\end{figure}

The lifetime mortality rates, $\mu(t)$, uniquely determine the survival statistics.\cite{Moore2016-rh} In Fig.~\ref{fig4}a we present the death age distributions for several disease parameter values. The disease has two lifespan-shortening effects: a short-term, acute effect that increases mortality during the disease, reducing lifespan by $\Delta t_{short}$; and a long-term, chronic effect that shifts the death age distribution to younger ages, further reducing lifespan by $\Delta t_{long}$. In Fig.~\ref{fig4}b we numerically calculate the ratio of acute to chronic effects. As with the GNM, we see that long-term effects dominate for younger individuals whereas short-term effects dominate for older individuals, and are essentially independent of disease severity $m \tau$.

We can also obtain simpler expressions for mortality effects -- particularly in the `weak' limit of small $m$ and $\tau$. These are useful  to develop an understanding of the origins of the effects exhibited by diseases in the GNM. 

\subsection*{Long-term effects}
While short-term survival mediates long-term effects, this coupling is small in the weak limit. For simplicity, here we will condition on short-term survival -- i.e. assume that individuals are alive at $t_{end}=t_{on}+\tau$ with excess damage $\Delta f$. 

Since mortality is determined by health, then the addition of exogenous damage $\Delta f$ at $t_{end}$ effectively ages an individual by $\Delta t_{long}$ where $f_0(t_{end}+\Delta t_{long})=f_0(t_{end})+\Delta f$. This is independent of the form of the mortality law. We obtain 
\begin{align}
 && \Delta t_{long} = \frac{1}{\alpha} \ln(1+\frac{\Delta f}{f_0(t_{end})}).
\label{eq:dtf}
\end{align}
This expression neglects a monotonic memory term which is small for young $t_{on}$, but that significantly decreases $\Delta t_{long}$ at old $t_{on}$ (Supplemental Eqn.~S75). 
Note that $\Delta t_{long}$ estimates the increase in biological age following disease.\cite{Mitnitski:2015} Using Eqn.~\ref{eq:deltaf}, and assuming small severities $m$  we obtain $\Delta t_{long} \approx \Delta f/(\alpha f_0(t_{end}))$. Further assuming small durations $\tau$ we obtain
\begin{align}
 && \Delta t_{long} \approx \frac{m \tau}{f_0(t_{on})}   (r + \frac{1-r}{\alpha \tau}).
\label{eq:smalldtf}
\end{align}
Since mortality only depends on $f$, $\Delta t_{long}$ estimates the long-term reduction in lifespan \emph{after} the survival of mild diseases -- excluding any short-term mortality during the disease. Since $f(t)$ increases with age, $\Delta t_{long}$ is largest in the youngest individuals -- independent of disease parameters $m$, $\tau$, and $r$.  For imperfect resilience, with $r<1$, chronic effects typically dominate the long-term impact of disease-survivors and  $\Delta t_{long} \approx m (1-r)/\left[ \alpha f_0(t_{on}) \right]$; these chronic effects are independent of $\tau$. We observed that COVID-19 has $r < 1$ whereas seasonal flu does not (see below).

\subsection*{Short-term effects}
We can use the hazard $\mu(t)$ in Eqn.~\ref{eq:hazard}
to solve for the survival probability $S(t)$, using $dS/dt = - \mu S$ (details are in the supplemental). Conditional on being alive $S=1$ at $t_{on}$ we obtain 
\begin{align}
&& S(t) = \exp[ - \frac{b}{\beta} (f_{on}/a)^{\beta/\alpha}(e^{\beta (t-t_{on})}-1)],
\label{eq:survival}
\end{align}
where $f_{on}$ is the frailty at $t_{on}$. The probability of mortality by the end of an acute disease is $1-S(t_{end})$ therefore we obtain the excess short-term mortality $\Delta p_{death}$ due to the acute disease by the difference in the survival function between using $f_{on}=f_0(t_{on})$ and $f_{on}+m$ at $t_{on}$. For small $m$ and $\tau$ we obtain 
\begin{align}
&&\Delta p_{death} \approx \frac{m\tau\beta}{\alpha}\frac{\mu_0}{f_{on}} = \frac{m\tau\beta}{\alpha}\frac{be^{\beta t_{on}}}{ae^{\alpha t_{on}}}.
\end{align}
We see that $\Delta p_{death} \propto e^{(\beta-\alpha)t_{on}}$ is highest for older individuals since $\beta > \alpha$. This is consistent with the observation of increasing short-term mortality with age in many  diseases. 

\subsection*{Comparing short- and long-term effects}
To compare short- and long-term effects, we need to estimate the years of life lost due to death during the disease -- all within the small $m$ and $\tau$ limit. We can approximate the remaining lifespan $\Delta t_D$ from the survival curve by imposing $S(t_{on}+\Delta t_D)=1/e$, this approximates the survival curve as a step function. Using Eqn.~\ref{eq:survival} we obtain $\Delta t_D = \beta^{-1} \ln(1+\beta/\mu_0(t_{on}))$. The years of life lost during acute disease is then $\Delta t_{short} = \Delta p_{death}\, \Delta t_D$ which gives
\begin{align}
 && \Delta t_{short} \approx \frac{m \tau}{f_{on}}   \frac{\mu_0}{\alpha} \ln(1+\frac{\beta}{\mu_0}).
\label{eq:duringdt}
\end{align}
In the limit of small $m$ and $\tau$, the ratio of short to long-term lifespan effects is then
\begin{align}
 && \frac{\Delta t_{short}}{\Delta t_{long}} \approx \frac{\beta}{\alpha} \ln(1+\frac{\beta}{\mu_0})/(\beta/\mu_0),
\label{eq:ratio}
\end{align}
where we have also allowed for maximal recovery after the disease ($r=1$). Interestingly, this ratio is independent of disease details. We note that $\ln(1+x)/x \approx 1$ for $x \approx 0$ and monotonically decreases towards $0$ with increasing $x = \beta/\mu$, i.e.\ with decreasing age. At large ages $\Delta t_{short}/\Delta t_{long} \approx \beta/\alpha >1$, so that short term mortality during disease affects lifespan more than long-term effects. Conversely, at sufficiently young ages, we expect long-term mortality effects after the disease to have greater impact on lifespan than short-term mortality during the disease. From our estimates of $\alpha$ and $\beta$, $\Delta t_{short}/\Delta t_{long}=1$ for $\mu \approx 0.024$. From all-causes mortality statistics from the U.S. population (Fig.~\ref{fig2}a, grey squares) we have $\mu \lesssim 0.024$ for ages $t_{on} \lesssim 70$, implying that $\Delta t_{short} < \Delta t_{long}$ for onset ages $< 70$. So, our phenomenological model indicates that most people would have a greater reduction of lifespan due to premature death long after the disease than from death during the disease. Similar results are observed away from the small $m$ and $\tau$ limit (see Fig.~\ref{fig4}) and in the GNM (see Fig.~\ref{fig3}a).

\subsection*{Long-term excess relative risk (ERR) and the Life-Span Study (LSS) of Atom-bomb survivors}
The Life-Span Study (LSS) of approximately 120,000 survivors of the atomic bombs dropped on Nagasaki and Hiroshima has tracked excess lifetime mortality due to radiation exposure for more than $50~\text{years}$, and found that excess relative risk decreased with age of exposure and was approximately linear with dosage.\cite{Preston:2003, Ozasa:2012} Deaths due to solid-tumor cancer predominate the excess mortality. 

Our phenomenological model allows for any source of exogenous damage $m$, not just from  disease. We recast it in terms of excess long-term hazard to be able to directly compare with the LSS analysis. Using Eqn.~\ref{eq:hazard} with $\tau=0$ we obtain
\begin{align}
    &&\mu(t) &= b e^{\beta t} \big( 1 + \frac{\Delta f}{a} e^{-\alpha t_{on}}\big)^{\beta/\alpha}.
    \label{eq:nonlin}
\end{align}
If we linearize in the hazard in $\Delta f$ we obtain
\begin{align}
&&    \mu(t) \approx b e^{\beta t}\big( 1 + \frac{\beta}{\alpha}\frac{\Delta f}{a}e^{-\alpha t_{on}} \big) = \tilde{\mu}_0(t,\vec{c}) \big( 1 + \gamma(\vec{c}) d e^{\theta t_{on}} \big) \label{eq:muapp},
\end{align}
where on the right we show a model of excess relative risk (ERR) from the LSS \cite{Preston:2003} -- here the covariates $\vec{c}$ such as sex, city, and birth year are indicated ($\text{ERR}\equiv \gamma(\vec{c}) d e^{\theta t_{on}}$). Qualitatively both the LSS and our approach have excess absolute risk\cite{Preston:2003} declining with age of exposure $t_{on}$ and with linear dose-response ($\Delta f$ or $d$ in Sv). We can identify $\theta = -\alpha$. Their model estimates $\alpha = 0.045$ (90\% CI: [0.031, 0.060]),\cite{Preston:2003} which is consistent with our estimate of $0.031$. We suggest that the increased radiation sensitivity at younger exposure ages reported by the LSS \cite{Preston:2003} may be a general effect of increased damage sensitivity at younger exposure ages.

Our phenomenological model also suggests different risk models that could be used with LSS data, such as including nonlinear effects with Eqn.~\ref{eq:nonlin}. Using $\alpha=-\theta$ and $\beta=0.089$ (Fig.~\ref{fig2}a), we estimate $\Delta f / a = 0.98 d$, where $d$ is the exposure dose in Sieverts (Sv).\cite{Preston:2003} This implies that the dose and the propagated damage $\Delta f$ are approximately equal, when expressed in natural units. Since survivable doses range up to $5$ Sv, the linearized approximation may be worse for younger individuals.

\subsection*{Parameterizations of COVID-19, influenza and Ebola}
Using published IFRs we estimated disease severity, $m$, for COVID-19,\cite{COVID-19_Forecasting_Team2022-ko}, influenza\cite{Lees2020-ze} and Ebola,\cite{Agua-Agum2015-qg} Table~\ref{tab:validation}. Studies of both COVID-19\cite{Muller2022-ze} and influenza\cite{Lees2020-ze} recorded health in terms pre-disease vs post-recovery frailty, $\Delta f$. This allowed us to estimate the resilience parameter for those diseases, $r$. Each column of Table~\ref{tab:validation} includes parameter estimates taken from the literature for populations at particular ages, including $\tau$, IFR, and $\Delta f$, together with our phenomenological model estimates for $m$ and $r$ using Supplemental Eqns.~S2 and S3, respectively (where possible). Observe that resilience was not significantly different from $1$ for influenza, but resilience was significantly lower for COVID-19. This may explain why COVID-19 is observed to have large long-term chronic effects \cite{Thompson:2022, Mulberry2021-mq}: Eqn.~\ref{eq:smalldtf} predicts that $r<1$ effects will dominate the chronic disease effects. See supplemental for details.

Disease severity, $m$, depends on individual robustness -- and is used to set the scale for both IFR and $\Delta f$. Note that while $m>1$, we observe physiologically reasonable $\Delta f \ll 1$.  We observed that as individuals age, their robustness follows a U-shaped curve: increasing from infancy to adulthood and then decreasing with advanced age (Supplemental Fig.~S5). In the case of COVID-19, this decreasing robustness with adult age paralleled the expected changes to frailty, $f$, suggesting a loss of robustness with increasing frailty. Consistent with this, comorbidities both increase the frailty index \cite{Searle:2008} and are major risk factors for mortality due to COVID-19 \cite{Lu2020-fg}.

The frailty index includes both physical and mental deficits \cite{Searle:2008}. A large UK study found that individuals whom suffered from severe COVID-19 showed reduced cognitive impairment $\sim2$ years post-infection comparable to effectively aging $\sim10$~years \cite{Cheetham2023-io}. Using Eqn.~\ref{eq:dtf}  we can estimate a generic aging effect from our model. Our $\Delta f$ indicates an effective aging of $\Delta t_{long}=6$~years for a median-aged 57.5 year-old -- comparable with the observed cognitive aging \cite{Cheetham2023-io}.

\begin{table}[h]
\caption{Disease Parameter Estimates for Specific Ages (95\% CI)} \label{tab:validation}
    \centering
    \begin{tabular}{l l l l}
    \hline
         & COVID-19 & Influenza (hospitalized)  & Ebola \\ \hline
     Age & 65 & 80.1\phantom{00} (SD: 8.7) & 16--44 \\
     $\tau$ (days) & 12 & 16.8 & 15.8 \\
     IFR & 0.017 (0.012-0.027) & 0.12\phantom{00} (0.11-0.14) & 0.65 (0.64-0.67) \\
     $m$ & \textbf{1.1}\phantom{00} (0.9-1.4) & \textbf{2.1}\phantom{000} (2.0-2.2) & \textbf{5.74} (5.66-5.82) \\
     $\Delta f$ & \textbf{0.063} (0.046-0.081) & \textbf{0.0065} (0.0041-0.0089) & \phantom{000000}-- \\
     $r$ & \textbf{0.94}\phantom{0} (0.93-0.96) & \textbf{0.998}\phantom{0} (0.997-1.000) & \phantom{000000}-- \\ \hline
    \end{tabular}
\end{table}

\section*{Discussion} 
We have developed and explored a three-parameter model of generic acute disease, which is built upon a generic network model (GNM) of organismal aging (age of onset $t_{on}$, severity $m$, and duration $\tau$). We evaluated short-term mortality outcomes using the excess infection fatality rate (IFR) and long-term mortality outcomes using the average reduction in lifespan due to the disease ($\Delta t_{tot}$).  We found that while mortality during acute diseases is highest for older populations, the total reduction in lifespan is highest for younger populations. The majority of the years of life lost for younger populations are due to premature deaths later in life. Older populations have worse short-term outcomes because they have greater frailty $f$ (worse health), which leads to a greater likelihood of death during the disease. Younger populations lose more years of life both because there is more to lose and more time for propagated damage $\Delta f$ to impact mortality at the end of life. 

Our results are qualitatively consistent with higher short-term mortality for older populations as reported for many acute diseases, including COVID-19,\cite{Levin:2020} SARS and MERS,\cite{Lu2020-fg} influenza,\cite{Simonsen1998-gs,Thompson:2003} Ebola,\cite{Agua-Agum2015-qg} varicella (chickenpox),\cite{Boelle:2002, Gil:2004} and meningococcal disease.\cite{Wang:2019} While the 1918 (``Spanish'') flu pandemic had much higher than expected mortality for younger adults, this appears to be a special (non-generic) case\cite{Taubenberger2006} partially due to the effects of age-varying immunological history.\cite{Gagnon2013, Simonsen1998-gs} 

Long-term impacts due to post-acute sequelae (PAS) are common.\cite{Ahmed:2020, Ngai:2010, Wilson:2018, Chen:2017, Wensaas:2012, vanAalst:2017, Souza:2019, Al-Aly:2021, Thompson:2022, Hickie:2006} We predict that such post-acute effects should increase with acute severity $m$, in qualitative agreement with, e.g., studies of long-COVID.\cite{Tsampasian2023-bq} Similar severity dependence is seen in ICU (intensive care unit) survivors.\cite{Wunsch2010} Our disease model is essentially one of exogenous damage, and so should be more general than just acute disease. Long-term studies of hip-fracture survivors have shown significant excess relative risk that is approximately independent of attained age \cite{Haentjens:2010, Katsoulis:2017} in agreement with our simple phenomenological model (Eqn.~\ref{eq:nonlin}). Atomic bomb survivors provide a unique long-term dataset for exogenous damage due to radiation \cite{Preston:2003} -- with exposure ages ranging from $0-60$ and with more than 50 years of followup. In agreement with our findings, lifetime risks are greatest for younger exposure ages $t_{on}$. 

Aging individuals exhibit changing robustness (resistance to damage) and resilience (recovery from damage) -- typically both declining with age.\cite{Whitson:2016, Ukraintseva:2016, Farrell:2022} Disease frequency typically increases with age,\cite{Palmer:2018} consistent with declining robustness. Robustness and resilience can be considered individual and disease-specific parameters since, e.g., vaccinations or prior exposure increase robustness to infectious disease while, e.g., medical care can improve recovery. Robustness could affect the frequency and/or severity of disease for older individuals (e.g. $t_{on}$ and $m$). Resilience could affect recovery and duration ($r$ and $\tau$). Our results are for a fixed severity ($m$) so direct comparisons between ages require caution. Nevertheless, the ratio $\Delta t_{short}/\Delta t_{long}$ is conditioned on the disease occurring, and is largely independent of disease severity (Fig.~\ref{fig4}b). The observation that the lifespan impact of disease can be much worse than the acute impact of disease for younger individuals is therefore independent of robustness.

Our model explicitly includes resilience through $r$. Smaller resilience ($r$) should lead to larger $\Delta f$ and thus worse long-term effects. Since resilience is expected to decrease with age,\cite{Ukraintseva:2016, Farrell:2022} we would expect more long-term effects in older individuals. The result would be a smaller ratio of $\Delta t_{short}/\Delta t_{long}$ for older individuals. 

Our disease model has no explicit age dependent dynamics, so all effects occur via individual health. We expect that short-term mortality will be worse with either worse health or older ages. Consistent with this, the prognosis of disease generally worsens with a higher frailty index $f$.\cite{Rockwood:2019, Howlett:2021} Multiple concurrent diseases are expected to combine additively through $f$, although saturation or exclusion effects may occur for severe or overlapping multimorbidities, respectively. While our phenomenological model has no age effect for $\Delta f$ at a given $m$, our GNM exhibits increasing $\Delta f$ with age. Furthermore, we expect that declining robustness with age (or declining health) will lead to larger $m$ and so larger long-term health impacts ($\Delta f$). Such effects are observed. For example, disability following hospitalization increases more with age \cite{Covinsky2003}, and more following ICU admission with frailty \cite{Ferrante2018}. Frailty hinders recovery from influenza \cite{Lees2020-ze}. Age is a risk-factor associated with post-COVID-19 conditions.\cite{Thompson:2022, Tsampasian2023-bq}, and with PAS of chikungunya virus disease.\cite{vanAalst:2017} 

Consistent with this picture, we observed that our estimates for disease severity, $m$, increased with age. For COVID-19, $m$ increased exponentially with age: commensurate with $f$ and consistent with a loss of robustness with increasing frailty. Although we did not have data to estimate age-related changes to resilience, we did observe that the seasonal flu showed nearly perfect resilience whereas COVID-19 indicated incomplete recovery ($r<1$). This could help explain the prevalence of COVID-19 PAS.\cite{Thompson:2022, Mulberry2021-mq} Parameterizing additional specific diseases will facilitate future studies to investigate disease-specific effects on lifetime mortality. 

Most studies of post-acute mortality effects  only have a $w \lesssim 5$ yr observation window.  We found that $w \gtrsim 20$ yr is needed to observe the largest mortality impacts, which we predict occur for smaller onset ages. Larger observation windows $w$ are needed. For shorter $w \lesssim 20$ windows, general health measures such as the frailty index $f$ \cite{Howlett2021} can be used to assess excess damage $\Delta f$ due to the disease. The effective cognitive aging of approximately 10 years due to long COVID-19 \cite{Cheetham2023-io} is consistent with our generic estimates using Eqn.~\ref{eq:dtf}. The relative ease with which mental deficits can be measured may make them a convenient way to measure follow up health post-infection. 

Our GNM disease model is stochastic and exhibits considerable individual variability in e.g., excess post-acute damage $\Delta f$ (see Fig.~\ref{fig1}b). For real diseases, we expect additional variability in the acute severity ($m$). Our models are restricted to adults (with $t \gtrsim 20$), due to similar restrictions on the GNM, frailty $f$, and Gompertz's law.  We expect adult males to experience worse short-term mortality risk, including both acute and chronic effects, due to their higher baseline risk (Supplemental Fig.~S6b). This sex-effect is seen in parasite-associated mortality \cite{Owens2002} and most infectious diseases \cite{Owens2002,Klein2016-vc}. 

Our simple phenomenological theory shares with the full disease model our assumptions that residual damage and mortality are determined by health via $f$. Subject to these assumptions, the qualitative agreement of our models indicates the potential universality of our results. From the phenomenological theory we see the key role of the exponential growth rates of mortality and frailty, $\beta$ and $\alpha$ respectively. Empirically we have $\beta>\alpha$, so short-term excess IFR ($\Delta p_{death}$) grows with age.  Our phenomenological theory also indicates that post-survivor years of life lost $\Delta t_{long}$ is universally greatest for younger adults -- a consequence of $\alpha>0$. 

We infer universal aspects of disease through the effects of direct ($m$) and secondary damage ($\Delta f$) in an aging population. We find large long-term effects at young onset ages. Including such age-effects in epidemic models, such as for COVID-19,\cite{Bubar2021-kw, Mulberry2021-mq} would help us better understand and mitigate the impacts of disease on societies. Researchers typically ask if it is better to vaccinate the old to reduce direct risk, or vaccinate the young to reduce overall infection prevalence.\cite{Bubar2021-kw} Similarly, cost effectiveness of e.g. rotavirus vaccine \cite{Jit:2007} or allocation of COVID-19 vaccine \cite{Goldstein:2021} often only consider mortality during disease. Often neglected are the potential chronic effects due to propagated damage, which we find are worse for the young. Our results could have significant implications for how we prioritize medical interventions across age. Long-term observational studies of health and mortality after acute disease or exposure are needed to better capture lifetime disease impacts.  

\section*{Data availability}
The disease model code used to generate the data presented in this paper are available at \url{https://github.com/RebeccaTobin/DiseaseModel}. The data used for plots is available on request from A.R..

\bibliography{ref}  

\section*{Acknowledgements}
We thank Kenneth Rockwood for helpful discussions, and Spencer Farrell for help with GNM code. This work was supported by the Natural Sciences and Engineering Research Council of Canada (NSERC) with operating Grant RGPIN-2019-05888 (ADR), and with a USRA award (RT). We acknowledge the Digital Research Alliance of Canada (DRAC) for compute resources. 

\section*{Author contributions statement}
R.T. conducted the simulations and data analysis and drafted the manuscript. G.P. conducted analysis of the phenomenological model and drafted the supplement. All authors contributed to the design of the research, the data interpretation and the manuscript production. 

\section*{Competing interests}
The authors declare no competing interests.

\section*{Additional information}
{\bf Correspondence} and requests for materials should be directed to A.R.

\end{document}


\flushbottom
\maketitle

\section{Generic network model (GNM) additional results}
\begin{figure}[h]
    \centering
    \includegraphics[width=0.5\textwidth]{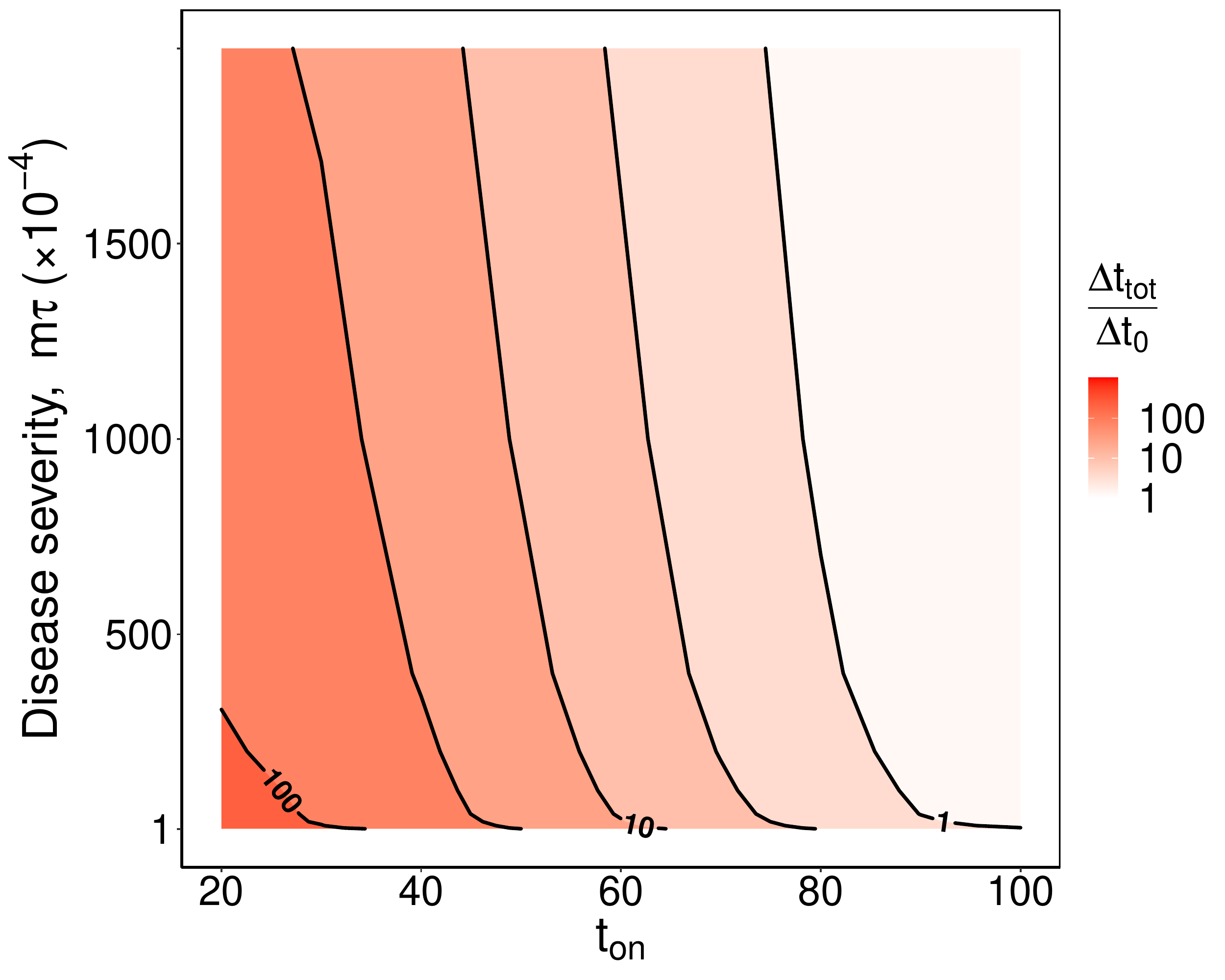}
    \caption{ {\bf Lifespan reduction ratio for different onset ages and disease severities. } Total years of life lost due to disease divided by life lost during only the acute phase. ($m$ = 0.02, $10^{-2} \leq \tau \leq 10^2$, $r = 1$)
    }
    \label{fig:gnmratio}
\end{figure}

Fig.~\ref{fig:gnmratio} shows the ratio of total years of life lost due to the disease, $\Delta t_{tot}$ divided by years of life lost during the disease, $\Delta t_0$, for a variety of onset times $t_{on}$ and disease severities ($m \tau$, with $m=0.02$). $\Delta t_{tot}$ includes all effects, whereas $\Delta t_0$ includes only the acute effects of the disease. Observe that $\Delta t_{tot}/\Delta t_{0}$ exceeds $100$ for younger individuals.

\begin{figure}[h] 
    \includegraphics[width=\textwidth]{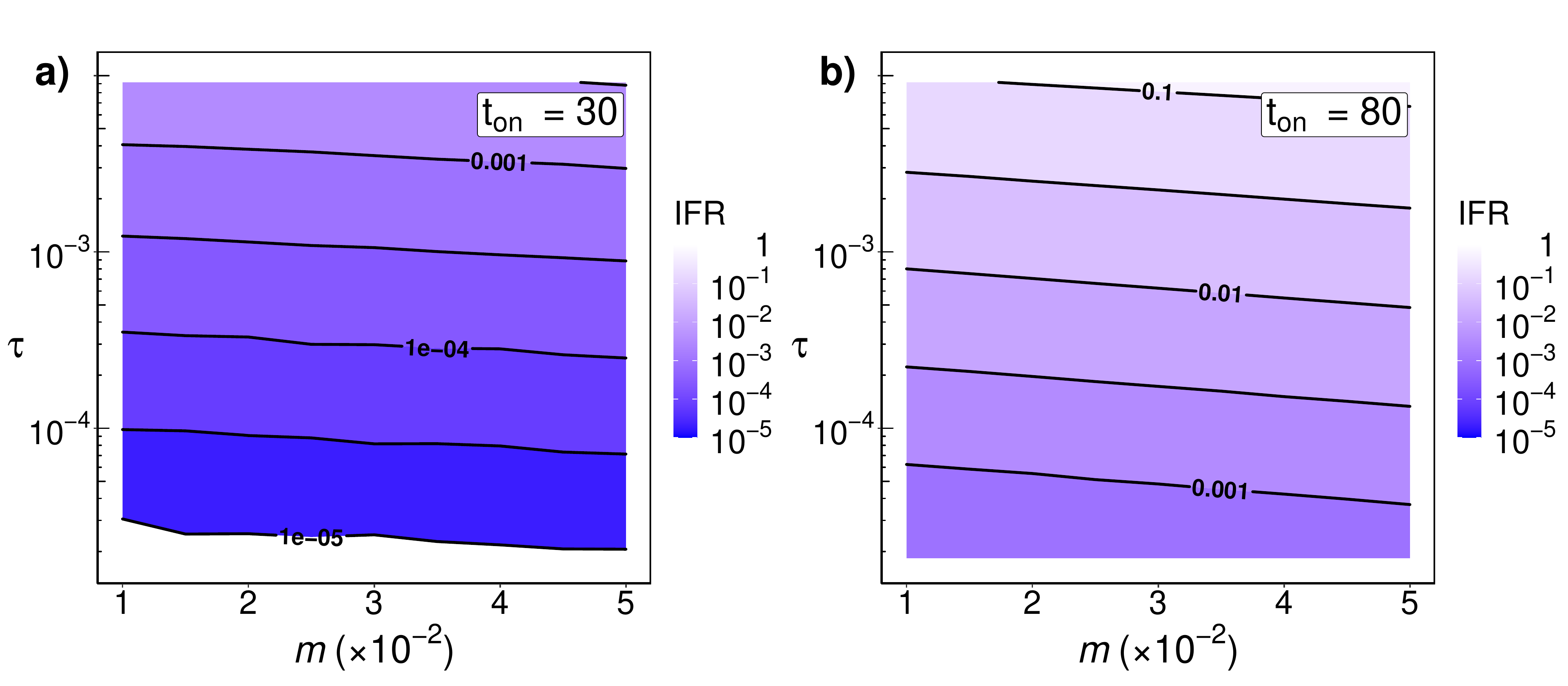} 
    \caption{\textbf{Infection fatality rate (IFR) as a function of $m$ and $\tau$.} IFR increases smoothly with increasing $m$ or $\tau$. {\bf(a)  30 year old.} {\bf (b) 80 year old.} Observe the much higher fatality rate for the 80 year old.
}
    \label{fig:ifr}
\end{figure}

\begin{figure}[h] 
    \includegraphics[width=\textwidth]{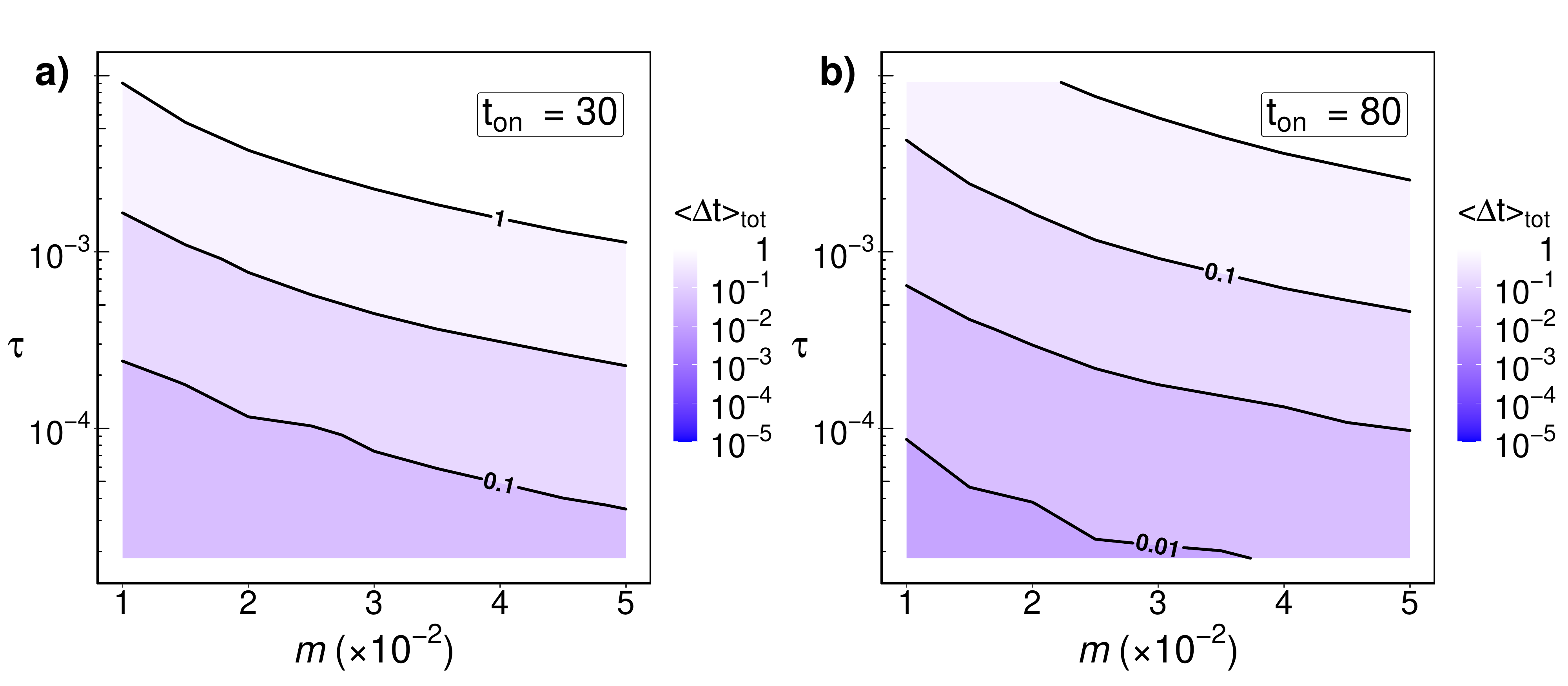} 
    \caption{\textbf{Total years of life lost as a function of $m$ and $\tau$.} Years of life lost increases smoothly with increasing $m$ or $\tau$. {\bf(a)  30 year old.} {\bf (b) 80 year old.} Whereas the older individuals died more often during disease, Figure~\ref{fig:ifr}, they saw a smaller net loss of life.
}
    \label{fig:dt}
\end{figure}

We performed a sensitivity analysis on the key GNM model parameters: $m$ and $\tau$. The effects of the two parameters were qualitatively similar, and increases to either $m$ or $\tau$ smoothly increased both infection fatality rate (IFR), Figure~\ref{fig:ifr}, and years of life lost, Figure~\ref{fig:dt}. In the main text we noted the important result that older individuals die more often during disease whereas younger individuals lose more total lifespan. This result holds when tuning either $\tau$ or $m$.

\FloatBarrier
\section{Validation for COVID-19, influenza, and ebola}
Here we provide details of fitting observed health and mortality data to our phenomenological model, and provide fits for selected diseases. Both the GNM and our phenomenological model are founded on two key hypotheses: (1) acute mortality is due to damage, and (2) this damage should cause secondary propagated damage. Secondary hypotheses are that robustness, via $m$, and resilience, via $r$, may vary by disease or due to individual risk factors (including health and age). While the main text primarily explores the key hypotheses, this supplemental section validates the secondary hypotheses using easily available data from influenza, COVID-19, and Ebola. Using acute mortality data we can estimate $m$ (and hence robustness effects). If individual health is followed post-recovery, we can also estimate $r$ (and hence resilience effects). 

Several studies have shown evidence of residual or collateral damage post-disease. Increased disability in activities of daily living \cite{Pizarro-Pennarolli2021-yy} and increased clinical frailty score \cite{Muller2022-ze} have both been observed after COVID-19 recovery. Similar effects are seen in other coronavirus': Middle East Respiratory Syndrome (MERS) and Severe Acute Respiratory Syndrome (SARS). SARS and MERS show long term deficits in fitness capacity and mental health -- including increased stress -- for up to a year post-recovery \cite{Ahmed:2020}. These deficits can lead to collateral, propagated damage due to the negative health effects of stress and dysfunction during the disease together with, e.g., the lack of positive effects of exercise. The main text deals with the consequences of this propagated damage. In general, residual damage may also be due to finite resilience $r<1$, i.e. acute damage that was not fully recovered from.

\begin{figure}[h] 
    \centering
    \includegraphics[width=.5\textwidth]{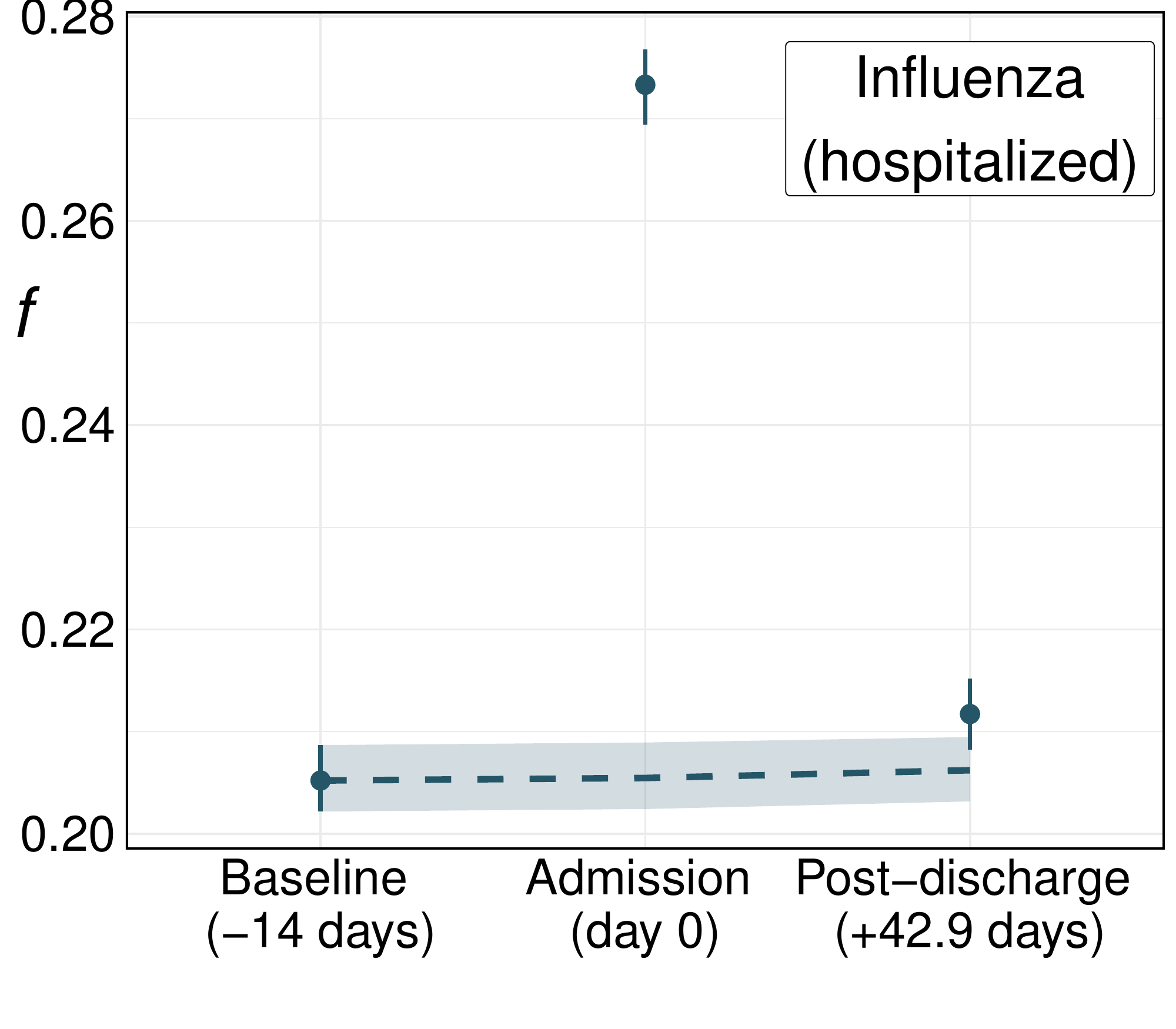}  
    \caption{\textbf{Changes to followup health due to influenza hospitalization.} Older individuals (average age 80) were measured for frailty index (FI, $f$) before, at and after hospitalization for influenza A or B. The band indicates the expected change to FI over the study period for the control group -- it is essentially constant. Data were extracted from survivor data in Fig.~2 of Lees \textit{et al}.\cite{Lees2020-ze} We approximated that individuals were in the hospital for 12.9 days -- this is the average length of stay for COVID-19 \cite{Faes2020-ra}, which is similar to influenza. \cite{Ludwig2021-le} Error bars are standard error in the mean.
}
    \label{fig:lees}
\end{figure}

Frailty has been observed to increase both after COVID-19 \cite{Muller2022-ze} and also after hospitalization due to influenza A/B \cite{Lees2020-ze}. In Figure~\ref{fig:lees} we present data from Lees \textit{et al's} study of hospitalizations due to confirmed influenza. They observed a marginally significant increase in the FI (frailty index $f$) post-influenza.  We use this influenza data from Lees \textit{et al}, and data from COVID-19 \cite{Muller2022-ze}, to estimate $r$ (resilience) and $m$ (including any robustness effects) for these diseases. We will also estimate $m$ from Ebola mortality data \cite{Agua-Agum2015-qg} -- where without health information we are unable to estimate $r$. 

The key statistic for estimating $m$ is the infection fatality rate (IFR). Estimating $r$ requires an additional estimate of $\Delta f$. $m$ represents the disease severity and is the fraction of damaged health attributes. $\Delta f$ represents the residual damage after some fraction, $r$, of the initial $m$ is repaired --- in addition to secondary propagated damage.

The IFR is simply the difference in survival between control and disease groups during the acute period,
\begin{align}
    \text{IFR} &\equiv \exp{\big[ -\int_{t_{on}}^{t_{on}+\tau} \mu(m=0,s)ds \big]} - \exp{\big[ -\int_{t_{on}}^{t_{on}+\tau}  \mu(m,s)ds \big]}\nonumber \\
    &\equiv S_c(t_{on},t_{on}+\tau)-S_d(t_{on},t_{on}+\tau) \label{eq:ifr}
\end{align}
where the survival of the control, $S_c$, and disease $S_d$, are defined by the corresponding terms in the preceding equation. The disease parameters are the age of onset, $t_{on}$, the disease strength, $m$, and duration $\tau$. While we do not know the forms of $S_c$ and $S_d$ from the GNM, we can compute analytic forms for the phenomenological model, see Section~\ref{sec:pheno} ($S_2$ using Eqns.~\ref{eq:hazard} and \ref{eq:s1s2s3}). This model assumes Gompertz' law and also that all mortality occurs due only to changes in the frailty index, $f$ (or ``FI''). 

For the phenomenological model we can algebraically invert Eqn.~\ref{eq:ifr} to yield the $m$-estimator,
\begin{align}
    m_{est} &= f_{on}\bigg( -\frac{\beta}{b}\frac{e^{-\beta (t_{on}+\tau)}}{1-e^{-\beta\tau}}\ln{\bigg[S_c(t_{on},t_{on}+\tau)-\text{IFR}\bigg]} \bigg)^{\alpha/\beta} - f_{on} \nonumber \\
    &= f_{on}\bigg( -\frac{\beta}{b}\frac{e^{-\beta (t_{on}+\tau)}}{1-e^{-\beta\tau}}\ln{\bigg[\exp{\bigg(-\frac{b}{\beta}e^{\beta(t_{on}+\tau)}(1-e^{-\beta\tau})\bigg)}-\text{IFR}\bigg]} \bigg)^{\alpha/\beta} - f_{on}, \label{eq:mest}
\end{align}
where $b$ and $\beta$ are Gompertz fit parameters from the healthy population, $\alpha\approx0.031$ is the FI growth exponent and $f_{on}$ is the control-group FI at the start of the disease.

\begin{figure}[h] 
    \includegraphics[width=.49\textwidth]{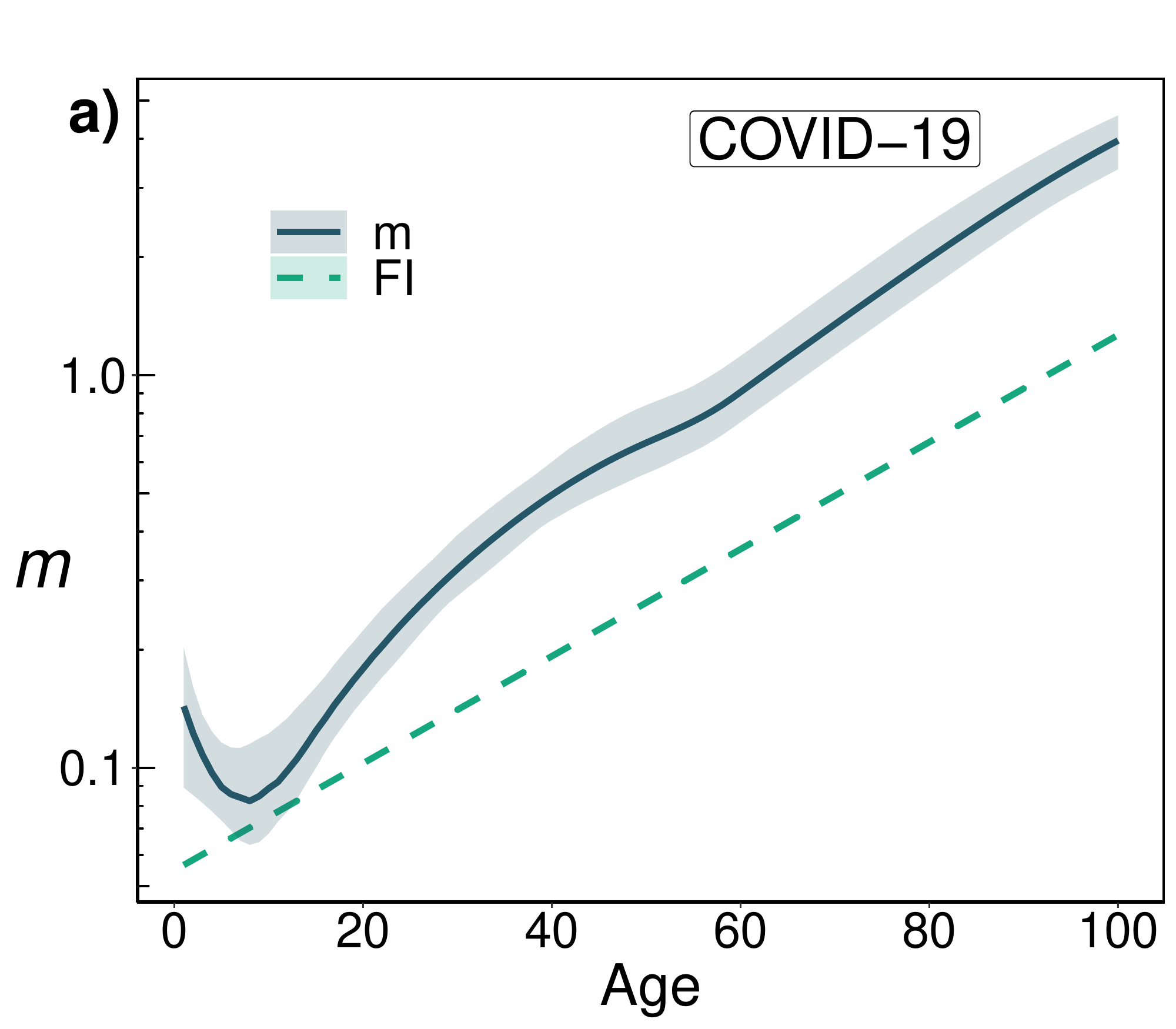} 
    \includegraphics[width=.49\textwidth]{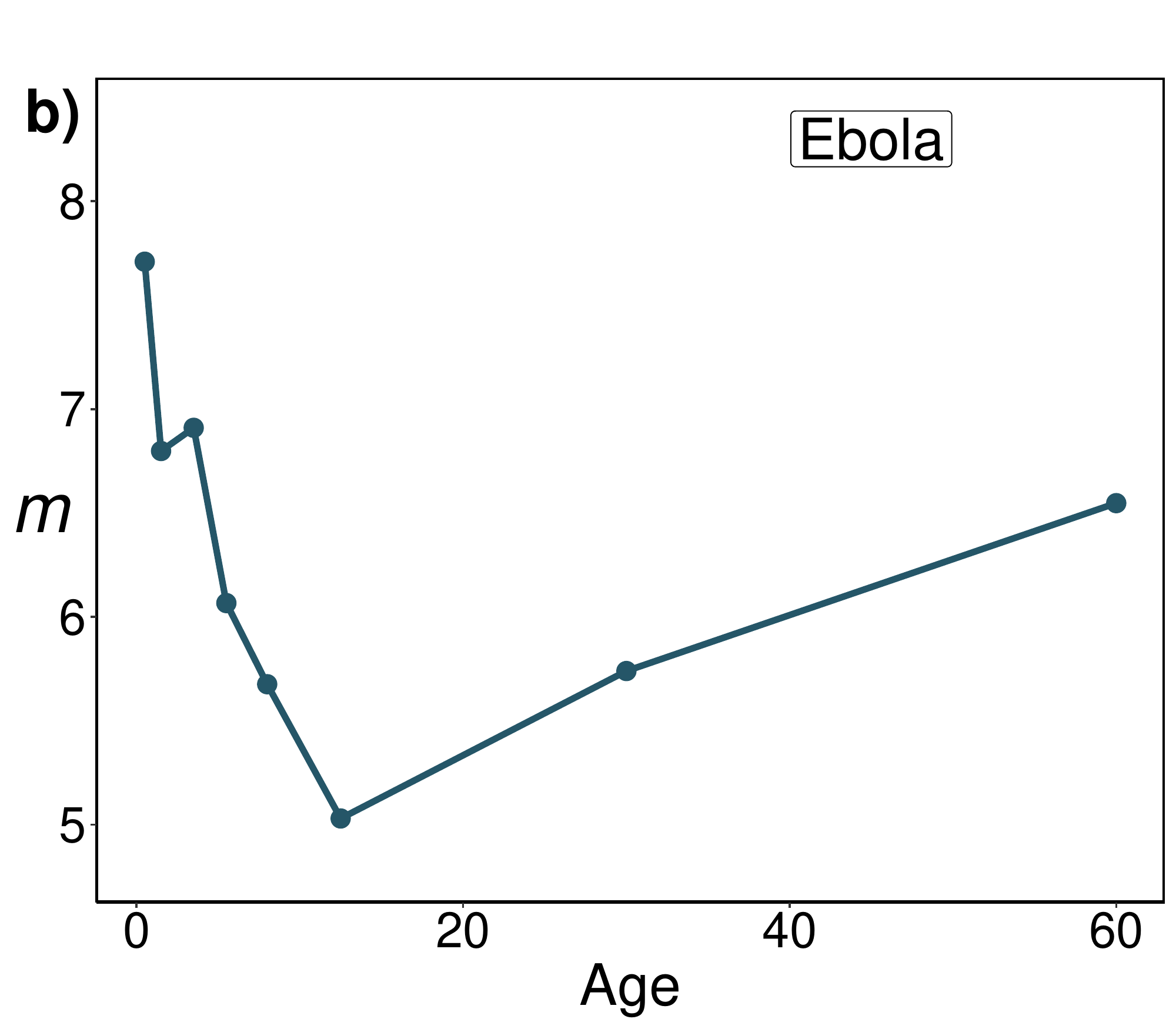} 
    \caption{\textbf{Disease intensity $m$ estimates, age-dependent.} Using infection fatality rate (IFR) data we  estimate $m$ as a function of age, which captures age-related changes to robustness (which decreases as $m$ increases). {\bf(a)  COVID-19.} We observed a strong age dependence for COVID-19.
    For adults, our estimates for $m$ increased exponentially with increasing age (blue solid line), paralleling the expected changes to the FI (green dashed line; same scale). This suggests that robustness may be a function of frailty. Source data: \textit{COVID-19 forecasting team, Lancet, 2022}\cite{COVID-19_Forecasting_Team2022-ko}.
      {\bf (b) Ebola.} We observed a small increase in susceptibility at young ages in both (note scales) COVID-19 (a)  and Ebola (b). For Ebola, the youngest group (< 1 years old) had higher mortality data than the oldest group (45+). Once individuals reach maturity, however, the disease effect, $m$, seems to be approximately constant with age for Ebola: with at most a linear age-dependence. Finally, note the much stronger effect of Ebola compared to COVID-19, with $m>1$ at all ages -- reflecting the much higher IFR. $f$ captures an individual's health state, so that for very strong diseases with $m>1$, acutely ill individuals are less healthy than the oldest healthy individual. Source data: \textit{Agua-Agum et al., N Engl J Med, 2015}\cite{Agua-Agum2015-qg}. The oldest age group (45+) was imputed as age 60 for convenience. 
}
    \label{fig:robustness}
\end{figure}

Using published IFR data we estimated $m$ for several diseases using Eqn.~\ref{eq:mest}. $m$ captures both the intrinsic severity of the disease and the individual's resistance to that disease i.e.\ robustness. Increases to $m$ with age reflect decreases to robustness (and vice versa). In Figure~\ref{fig:robustness} we present $m$ as a function of age for COVID-19 (a) and Ebola (b).  As we would expect, robustness increases from infancy to adulthood, causing $m$ to decrease in both diseases. Also expected is that robustness then decreases with increasing age during adulthood, causing $m$ to increase. This increase of $m$ is much faster with COVID-19 (note log-scale) than with Ebola. For COVID-19 the increase of $m$ (decrease of robustness) approximately parallels the increase of the FI (frailty, $f$). The different behavior of COVID-19 and Ebola supports the hypothesis that robustness is disease dependent. Note that these robustness effects are in addition to the age-effects of acute mortality discussed in the main paper with constant $m$.  The combination of age effects are qualitatively consistent with known mortality risk factors for coronavirus' such as COVID-19, including a strong age dependence and the magnifying effects of comorbidities -- which increase $f$ -- such as hypertension, diabetes and chronic lung disease \cite{Lu2020-fg}. 

%

After the disease ends, the residual damage depends on the resilience, $r$, as
\begin{align}
    \Delta f &= m(e^{\alpha \tau}  - r), \label{eq:df}
\end{align}
which permits us to estimate $r$ if both $m$ and $\Delta f$ are known. Conversely, if we know $r$ then we can estimate $\Delta f$.

Parameter estimates for three infectious diseases are reported in Table~\ref{tab:validation}. For influenza and COVID-19 we were able to estimate both the acute effect --- via IFR --- and the chronic effect via observed changes to patient FI, $f$. The chronic effect is mediated by the resilience parameter, $r$, which determines how much endogenous damage persists after the disease. For influenza, we observed that the resilience parameter is very close to $1$, suggesting near complete recovery. In contrast, COVID-19 showed far more residual damage with a notably smaller $r$ estimate. This qualitative difference is consistent with the many reported long term side effects for COVID-19, including long COVID \cite{Thompson:2022} and COVID complications \cite{Mulberry2021-mq}. Note that any $r<1$ can lead to large changes in residual damage $\Delta f$ -- compare $\Delta f_{obs}$ (observed) and $\Delta f_{r=1}$ (modeled with $r=1$) in Table~\ref{tab:validation}.

For COVID-19, we estimated $m$ using IFR from the COVID-19 Forecasting Team \cite{COVID-19_Forecasting_Team2022-ko} and compared it to $\Delta f_{obs}$ from a different study \cite{Muller2022-ze}. This latter study measured post-COVID recovery changes to the clinical frailty scale, which was converted to the frailty index $f$ using a linear model trained using data from a third study \cite{Moreno-Arino2020-qp}. In short, non-hospitalized COVID-19 patients saw an increase of $1$ in the clinical frailty scale which we estimated as a change of $\Delta f_{obs}=0.063$. For influenza, we extracted data from Lees \textit{et al.} which captured both mortality and $\Delta f_{obs}$ for hospitalized influenza A/B patients. The study population were draw from hospitalized patients, which explains why their IFR was so high.

\begin{table}[h]
\caption{Disease Parameter Estimates for Specific Ages (95\% CI)} \label{tab:validation}
    \centering
\begin{threeparttable}
    \begin{tabular}{l l l l}
    \hline
         & COVID-19 & Influenza (hospitalized) \cite{Lees2020-ze} & Ebola\cite{Agua-Agum2015-qg} \\ \hline
     Age & 65 & 80.1\phantom{000} (SD: 8.7) & 16--44\tnote{(1)} \\
     $\tau$ (days) & 12\tnote{(2)} & 16.8\tnote{(3)} & 15.8 \\
     IFR & 0.017\phantom{0} (0.012-0.027) & 0.12\phantom{00} (0.11-0.14) & 0.65\phantom{00} (0.64-0.67) \\
     $m_{est}$\tnote{(4)} & 1.1\phantom{000} (0.9-1.4)\tnote{(5)} & 2.1\phantom{000} (2.0-2.2) & 5.74\phantom{00} (5.66-5.82) \\
     $\Delta f_{obs}$ & \textbf{0.063}\phantom{0} (0.046-0.081)\tnote{(6)} & \textbf{0.0065} (0.0041-0.0089)\tnote{(7)} & -- \\
     $r_{est}$ & \textbf{0.94}\phantom{00} (0.93-0.96) & \textbf{0.998}\phantom{0} (0.997-1.000) & -- \\ 
     $\Delta f_{r=1}$\tnote{(8)} & 0.0011 (0.0010-0.0014) & 0.0030 (0.0028-0.0032) & 0.0078 (0.0077-0.0079) \\ \hline
    \end{tabular} 
\begin{tablenotes}
\item[(1)] Estimates are for age 30.
\item[(2)] Typical COVID-19 recovery time, 10--14~days \cite{medlineplus}.
\item[(3)] Symptom to hospitalization plus length of time in hospital (mean) for COVID-19 \cite{Faes2020-ra}, which is similar to influenza\cite{Ludwig2021-le}.
\item[(4)] $m_{est}$ is the estimated value for $m$ using Eqn.~\ref{eq:mest}.
\item[(5)] From fit to acute mortality data \cite{COVID-19_Forecasting_Team2022-ko}.
\item[(6)] Non-hospitalized COVID-19 patients showed a median 1 point increase in clinical frailty score \cite{Muller2022-ze}, which was converted to the FI scale using linear regression from a comparison study \cite{Moreno-Arino2020-qp}.
\item[(7)] Data were extracted from Fig.~2 of Lees \textit{et al.}\cite{Lees2020-ze}
\item[(8)] $\Delta f_{r=1}$ was computed using $m_{est}$ and Eqn.~\ref{eq:df} with $r=1$.
\end{tablenotes}
\end{threeparttable}
\end{table}

\FloatBarrier

\section{Phenomenological model} \label{sec:pheno}
In our generic network model (GNM) the frailty index $f$ is the average number of damaged nodes. Our model of disease increases the frailty index by $m$ at time $t_{on}$ by damaging $m N$ nodes at random (where $N=10^4$ is the number of nodes). At the end of the disease, a fraction, $r$, of the damaged nodes are repaired. The model is applied at age $t_{on}$ for duration $\tau$. Mortality occurs when the two most connected nodes of the GNM are simultaneously damaged. Damage promotes damage, thus strongly coupling mortality to frailty. A simple phenomenological model captures this effect and provides analytical expressions for the disease process, which are qualitatively consistent with the GNM.

The essential assumption of the phenomenological model is that the increase in mortality risk with age is solely due to increasing frailty. The mortality risk (hazard) is assumed to obey Gompertz' law,
\begin{align}
    \mu &= b e^{\beta t}
\end{align}
where $b$ and $\beta$ are estimated using all-causes mortality data of risk versus age (e.g.\ Fig.~2a). Frailty represents an individual's state of ill-health, it is equal to the average number of health deficits an individual has. The frailty index is known to increase exponentially \cite{Mitnitski:2015} as
\begin{align}
    f &= a e^{\alpha t}.
\end{align}
Thus the essential assumption of our model is that
\begin{align}
    \mu &= b (e^{\alpha t})^{\beta/\alpha} = b \bigg( \frac{f}{a} \bigg)^{\beta/\alpha}. \label{eq:mufi}
\end{align}

\subsection{Damage and resilience}
Our disease model is to impose exogenous damage. The effect of damaging a fraction of nodes, $m$, at time $t_{on}$ is to shift the frailty index (FI),
\begin{align}
    \lim_{\epsilon\to0^+} f(t_{on}+\epsilon) &= ae^{\alpha t_{on}} + m.
\end{align}
The individual then accumulates damage as
\begin{align}
    f(t_{on} < t < t_{on} +\tau) &= (ae^{\alpha t_{on}} + m)e^{\alpha t} \nonumber\\
    &= f(t,m=0) + m + m(e^{\alpha t}-1) \nonumber\\
    &= \text{Control FI} + \text{Exogenous FI} + \text{Propagated FI}.
\end{align}
Where the last two terms were caused by the disease.

Resilience is modelled by repairing a fraction, $r$ of the exogenous damage at the end of the disease. This leaves an effective `dose' of residual damage,
\begin{align}
    \Delta f &= m(e^{\alpha \tau}  - r)
\end{align}
that persists after the disease.

The resilience can be generalized to separately recover from either exogenous ($r$) or propagated ($r_{prop}$) damage using,
\begin{align}
    \Delta f_{gen}(r,r_{prop}) &= m(1-r) + m(e^{\alpha \tau}-1)(1-r_{prop}). \label{eq:mgen}
\end{align}
Note that $\Delta f_{gen}(r=r,r_{prop}=0)=\Delta f$. The generalized resilience modifies $\Delta f$ and therefore the post-disease mortality (via the hazard rate). The generalized resilience can be used to decouple the acute and chronic phases of the disease by arbitrarily tuning the effects of propagated damage, $m(e^{\alpha \tau}-1)$, against the effects of direct damage, $m$.  In this supplemental we take $r_{prop}=0$, but Eqn.~\ref{eq:mgen} can be used in any expression using $\Delta f$.

\subsection{Risk formalism}
We formalize risk with time-to-event statistics.\cite{Moore2016-rh} The death age distribution is defined as the time-to-event probability density function and is simply the product
\begin{align}
    p(t) &= \mu(t)S(t), \label{eq:ptmus}
\end{align}
where $\mu(t)$ is the risk (hazard) defined as the conditional probability of dying between $t$ and $t+dt$ (divided by $dt$), and $S(t)$ is the probability of surviving to time $t$.\cite{Moore2016-rh} The survival can be calculated using
\begin{align}
    S(t) &= \exp{\bigg( -\int_0^t \mu(u) du \bigg)}.
\end{align}
Observed that
\begin{align}
    p(t) &= -\frac{dS(t)}{dt}, \label{eq:pdsdt}
\end{align}
and conversely,
\begin{align}
    S(t)=1-\int_0^t p(u) du. \label{eq:sintp}    
\end{align}
The conditional distribution given that death occurs after some reference time, $t_r$, is
\begin{align}
    p(t,m | t > t_r) &= \label{eq:ptcond}
\begin{cases}
0 & t \leq t_r  \\
\frac{p(t,m)}{S(t_r,m)} & t > t_r 
\end{cases}
\end{align}
where $S(t_r,m)$ ensures normalization, $\int_0^\infty p(t > t_r,m) dt = 1$. The associated survival function is
\begin{align}
    S(t,m | t > t_r) &= \label{eq:scond}
\begin{cases}
1 & t \leq t_r  \\
\frac{S(t,m)}{S(t_r,m)} & t > t_r
\end{cases}
\end{align}
using $S(t)=1-\int_0^t p(t) dt$.

Finally, a useful result for later is to apply integration by parts using $p(t)=-dS/dt$ (Eqn.~\ref{eq:pdsdt}),
\begin{align}
    \int_a^b tp(t) dt &= -\int_a^b t\frac{dS}{dt} dt \nonumber \\
    &= aS(a)-bS(b) + \int_a^b S dt \label{eq:dtparts}.
\end{align}
Note that
\begin{align}
    \int_0^{\infty} tp(t) dt &= \int_0^{\infty} S dt. \label{eq:tptos}
\end{align}

\subsection{Disease} 
The effects of the disease on survival can be formalized by calculating the FI, $f$, as a function of disease parameters. Using Eqn.~\ref{eq:mufi} we can then calculate the mortality risk and therefore the survival using the risk formalism. \cite{Moore2016-rh} The disease parameters are
\begin{itemize}
    \item the severity, $m$, equal to the increase in FI during the disease,
    \item the duration, $\tau$,
    \item the age of onset, $t_{on}$, and
    \item the resilience, $r$, equal to the fraction of $m$ that is repaired at the end of the disease.
\end{itemize}
For convenience, we define
\begin{align}
    t_{end} \equiv t_{on} + \tau
\end{align}
as the end time of the disease. The control is the special case with $m=0$.

We can track $f$ because we know how much damage we're adding and therefore we know,
\begin{align}
    f(t) &= 
    \begin{cases}
        ae^{\alpha t} &t < t_{on} \\
        (ae^{\alpha t_{on}} + m)e^{\alpha (t-t_{on})} = (f(t_{on})+m)e^{\alpha (t-t_{on})} = (a+me^{-\alpha t_{on}})e^{\alpha t} &t_{on} \leq t < t_{end} \\ 
        ((f(t_{on})+m)e^{\alpha (t_{end}-t_{on})} - rm)e^{\alpha (t-t_{end})}  = f(t,m=0)+\Delta fe^{\alpha (t-t_{end})}  = (a+\Delta fe^{-\alpha t_{end}})e^{\alpha t} &t_{end} \leq t 
    \end{cases}
\end{align}
where $\Delta f\equiv m(e^{\alpha\tau}-r)$ and $t_{end}\equiv t_{on}+\tau$. Observe that the disease is equivalent to introducing some initial damage, thus increasing the FI by $m$. Conversely, we can view adding initial damage as aging the individual,
\begin{align}
    f(t) &= ae^{\alpha (t+\delta)} \label{eq:delta0}
\end{align}
for gained age, $\delta$, defined by Eqn.~\ref{eq:delta0}. See Section~\ref{sec:ba} for details.

We can then calculate the hazard using Eqn.~\ref{eq:mufi} to yield,
\begin{align}
    \mu(t) &= 
    \begin{cases}
    b \big(\frac{f(t,m)}{f(t,m=0)}\big)^{\beta/\alpha}e^{\beta t} &t < t_{on} \\
    b \bigg( \frac{(f(t_{on})+m)e^{-\alpha t_{on}}}{a}\bigg)^{\beta/\alpha}e^{\beta t} = b \big(\frac{f(t,m)}{f(t,m=0)}\big)^{\beta/\alpha}e^{\beta t}  &t_{on} \leq t < t_{end} \\ 
    b \bigg( \frac{(f(t_{end},m=0)+\Delta f)e^{-\alpha t_{end}}}{a}\bigg)^{\beta/\alpha}e^{\beta t} = b \big(\frac{f(t,m)}{f(t,m=0)}\big)^{\beta/\alpha}e^{\beta t}  &t_{end} \leq t. \label{eq:hazard}
    \end{cases}
\end{align}
where $f(t,m)/f(t,m=0) = f(t,m)\exp{(-\alpha t)}/a$ is the relative FI of the case vs control. Observe that the ratio $f(t,m)/f(t,m=0)$ has only step-function time-dependence and is piecewise constant: before, during and after the disease.

Using the fact that $f(t,m)/f(t,m=0)$ is piecewise-constant we can easily compute the survival function by breaking up the integral as follows:
\begin{align}
    S(t) &\equiv \exp{\bigg(-\int_0^t \mu(u) du \bigg)} \nonumber \\
    &= \exp{\bigg(-\int_0^{\text{min}(t,t_{on})} \mu(u) du \bigg)}\exp{\bigg(-I( t > t_{on})\int_{t_{on}}^{\text{min}(t,t_{end})} \mu(u) du \bigg)} \exp{\bigg(-I(t > t_{end})\int_{t_{end}}^t \mu(u) du \bigg)} \nonumber \\
    &= S_1(\text{min}(t,t_{on}))S_2(\text{min}(t,t_{end}))S_3(t) \label{eq:s1s2s3}
\end{align}
where $I(x)$ is the indicator function which is $1$ for true and $0$ otherwise. Observe that $S$ can be split up into a product of optional terms $S_1$, $S_2$ and $S_3$,
\begin{align}
    S(t) &= 
    \begin{cases}
    S_1(t) &t \leq t_{on} \\     
    S_1(t_{on})S_2(t) &t_{on} \leq t \leq t_{end} \\ 
    S_1(t_{on})S_2(t_{end})S_3(t)  &t_{end} \leq t. \label{eq:s1s2s3b}
    \end{cases}
\end{align}
Note that $S_1(t)$ is the probability of survival before the disease, $S_2(t)$ is the conditional probability of survival during the disease, and $S_3(t)$ is the conditional probability of survival after the disease. Conditioning on surviving to the start of the disease is easily achieved by setting $S_1(t_{end})=1$.

\textbf{Proof}\\
\begin{align}
    S_1(t \leq t_{on}) &\equiv \exp{\bigg(-\int_0^t b e^{\beta u} du \bigg)} = \exp{\bigg(\frac{b}{\beta}(1 - e^{\beta t})\bigg)}  = \exp{\bigg(-\frac{\mu(t)}{\beta}(1-e^{-\beta t})\bigg)}\\
    S_2(t_{on} < t \leq t_{end}) &\equiv \exp{\bigg(-\int_{t_{on}}^t b \bigg( \frac{(f(t_{on})+m)e^{-\alpha t_{on}}}{a}\bigg)^{\beta/\alpha} e^{\beta u} du \bigg)} = \exp{\bigg(-\frac{\mu(t)}{\beta}(1-e^{-\beta (t-t_{on})})\bigg)}\\
    S_3(t \geq t_{end}) &\equiv \exp{\bigg(-\int_{t_{end}}^t b \bigg( \frac{(f(t_{end},m=0)+\Delta f)e^{-\alpha t_{end}}}{a}\bigg)^{\beta/\alpha}\exp{(\beta u)} du \bigg)} = \exp{\bigg(-\frac{\mu(t)}{\beta}(1-e^{-\beta (t-t_{end})})\bigg)}.
\end{align}
then
\begin{align}
    S(t) &= S_1(t) &t \leq t_{on} \\     
    S(t) &= S_1(t_{on})S_2(t) &t_{on} \leq t \leq t_{end} \\ 
    S(t) &= S_1(t_{on})S_2(t_{end})S_3(t)  &t_{end} \leq t. 
\end{align}

\textbf{QED.}

Note that the control survival is simply
\begin{align}
    S(t,m=0) = S_1(t).
\end{align}

\subsection{Timescales} \label{sec:timescales}
Consider the conditional probability of surviving from a reference time $t_r$ to time $t$,
\begin{align}
    S_r \equiv \exp{\bigg(-\int_{t_r}^t \mu(u) du\bigg)}.
\end{align}
It will prove convenient to deal with the characteristic timescale of $S_r$, that is the time it takes $S_r$ to decay to a threshold value. A characteristic timescale, $t_d$, is defined by the condition
\begin{align}
    S_n(t_{d}) &= \frac{1}{d}.
\end{align}
For example, $d=2$ gives the halflife.

For the control case, $m=0$, the general solution is
\begin{align}
    t_{d} &= \frac{1}{\beta}\ln{\bigg(\frac{\beta}{b}\ln{(d)}+e^{\beta t_r}\bigg)}.
\end{align}
Observe that $t_d > t_r$.

We note the exponential timescale (with $d=e$) is
\begin{align}
    t_{e} &\equiv  \frac{1}{\beta}\ln{\bigg(\frac{\beta}{b}+e^{\beta t_r}\bigg)}    \label{eq:te}
\end{align}
for reference time $t_r$ ($t_r=0$ for $S$ and $S_1$, $t_r = t_{on}$ for $S_2$ and $t_r = t_{end}$ for $S_3$).

We will encounter integrals of the form
\begin{align}
    \int_{t_r}^\infty S(u) du.
\end{align}
For $t_r=0$ this is the average survival time.\cite{Moore2016-rh} We will find that the integral is generally a double exponential of the form $\exp{(\exp{(t)})}$ and hence is not analytically integrable. We find that a reasonable approximation is
\begin{align}
    \int_{t_r}^\infty S(u) du &\approx \int_{t_r}^{t_d} 1 du = t_d - t_r, \label{eq:sint}
\end{align}
i.e. that the survival function is approximately a step function equal to $1$ until some characteristic time, $t_d$, then $0$ afterwards. The approximation is good for survival functions that drop rapidly, which is achieved by a large $\mu$ e.g.\ $S_3$ for old individuals.

\subsection{Effective Age} \label{sec:ba}
Our phenomenological model maps time to damage and vice versa. Hence  exogenous damage introduced by the disease has the same effect as aging some period, $\delta$. Hence, we can think of an individual of age $t$ as having \textit{biological age} $t+\delta$, which represents their state of health i.e.\ they are as healthy as a control individual of age $t+\delta$.

Consider the hazard after the end of the disease,
\begin{align}
    \mu(t > t_{end}) &= b \bigg( \frac{(f(t_{end},m=0)+\Delta f)e^{-\alpha t_{end}}}{a}\bigg)^{\beta/\alpha}e^{\beta t}
\end{align}
using Eqn.~\ref{eq:hazard}. Observe that we can write
\begin{align}
    \mu(t > t_{end}) &= b e^{\beta\delta}e^{\beta t} = b e^{\beta (t+\delta)} = \mu(t+\delta,m=0). \label{eq:ba}
\end{align}
That is, the hazard of the disease is equal to the hazard of the control, shifted by some effective gain in age, $\delta$. Doing the algebra we have
\begin{align}
    \delta &= \frac{1}{\alpha}\ln{\bigg(1 + \frac{\Delta fe^{-\alpha t_{end}}}{a} \bigg)} = \frac{1}{\alpha}\ln{\bigg(1 + \frac{\Delta f}{f_{end}} \bigg)}  \label{eq:delta}
\end{align}
where $f_{end}$ is the control FI at time $t_{end}$.

\subsection{Lost lifespan}
Exposure to disease will cause a loss of lifespan compared to control. The simplest non-trivial model is to compare the expected loss of lifespan,
\begin{align}
    \langle \Delta t \rangle &\equiv \int_0^\infty t p(t,m=0) dt - \int_0^\infty t p(t,m) dt \nonumber \\
    &= \int_0^\infty S(t,m=0) - S(t,m) dt \label{eq:dt}
\end{align}
where the last line comes Eqn.~\ref{eq:tptos}, and $\langle \Delta t \rangle$ is the expected number of lost years of life, $p(t,m)$ is the death age distribution, $S(t,m)$ is the survival function, and $m=0$ denotes the control group. If we consider only individuals whom lived long enough to get the disease then Eqn.~\ref{eq:dt} is equivalent to the negative of the excess lifetime risk.\cite{Thomas1992-lt} We leave the expression in its most general form and later set $S_1(t_{on})=1$ to condition on surviving to the start of the disease. 

We are particularly interested in the acute versus chronic effects of the disease. We define acute/short and chronic/long diseases by splitting the effects of the disease into time intervals $0 \leq t < t_{end}$ (acute/short) and $t \geq t_{end}$ (chronic/long). We derive the expressions in the Sections~\ref{sec:shortproof} and \ref{sec:longproof}. In short, the key assumptions are (1) the acute/short phase has identical statistics to the disease until $t_{end} = t_{on}+\tau$ and then reverts to the control and (2) conversely, the chronic/long phase has identical statistics to the control until $t_{end}$ then obeys disease statistics. The survival function is causal, $S(t)$ integrates from $0$ to $t$, never past $t$. The hazard has no memory, it sees only the current $f$ and $t$. Hence the statistics are also causal since $p(t)=\mu(t)S(t)$. This has implications for normalizing the short and long diseases.

The short (acute) disease distribution is
\begin{align}
    p_{short}(t,m) = 
\begin{cases}
p(t,m) & t < t_{end} \\
p(t,m=0)\frac{S(t_{end},m)}{S(t_{end},m=0)} & t \geq t_{end}. \label{eq:pshort}
\end{cases}
\end{align}
This ensures that the survival is the same up until the end of the disease for both the acute and full diseases. The ratio of survival functions ensures that $p_{short}(t,m)$ is properly normalized (i.e.\ $\int_0^\infty p_{short}(t,m)dt=1$). This can be equivalently written in terms of $S_1$, $S_2$ and $S_3$ as
\begin{align}
S_{1,short}(t,m) &= S_{1}(t,m=0) = S_1(t,m) \nonumber \\
S_{2,short}(t,m) &= S_2(t,m) \nonumber \\
S_{3,short}(t,m) &= S_{3}(t,m=0),
\end{align}
where $S_1$, $S_2$ and $S_3$ are defined above (Eqn.~\ref{eq:s1s2s3}). That is, the survival curve differs from control only during the acute phase of the disease. See Section~\ref{sec:shortproof} for proof.

Similarly the long distribution is,
\begin{align}
    p_{long}(t,m) = 
\begin{cases}
p(t,m=0) & t < t_{end}  \\
p(t,m)\frac{S(t_{end},m=0)}{S(t_{end},m)} & t \geq t_{end}. \label{eq:plong}
\end{cases}
\end{align}
This ensures that the survival is the same up until the end of the control for both the chronic and control. The ratio of survival functions ensures that $p_{short}(t,m)$ is properly normalized. This can be equivalently written in terms of $S_1$, $S_2$ and $S_3$ as
\begin{align}
S_{1,long}(t,m) &= S_{1}(t,m=0) = S_1(t,m) \nonumber \\
S_{2,long}(t,m) &= S_2(t,m=0) \nonumber \\
S_{3,long}(t,m) &= S_{3}(t,m).
\end{align}
where $S_1$, $S_2$ and $S_3$ are defined above (Eqn.~\ref{eq:s1s2s3}). That is, the survival curve differs from control only during the chronic phase of the disease (after the disease). See Section~\ref{sec:longproof} for proof. 

\begin{figure}[h] 
    \centering
    \includegraphics[width=0.49\textwidth]{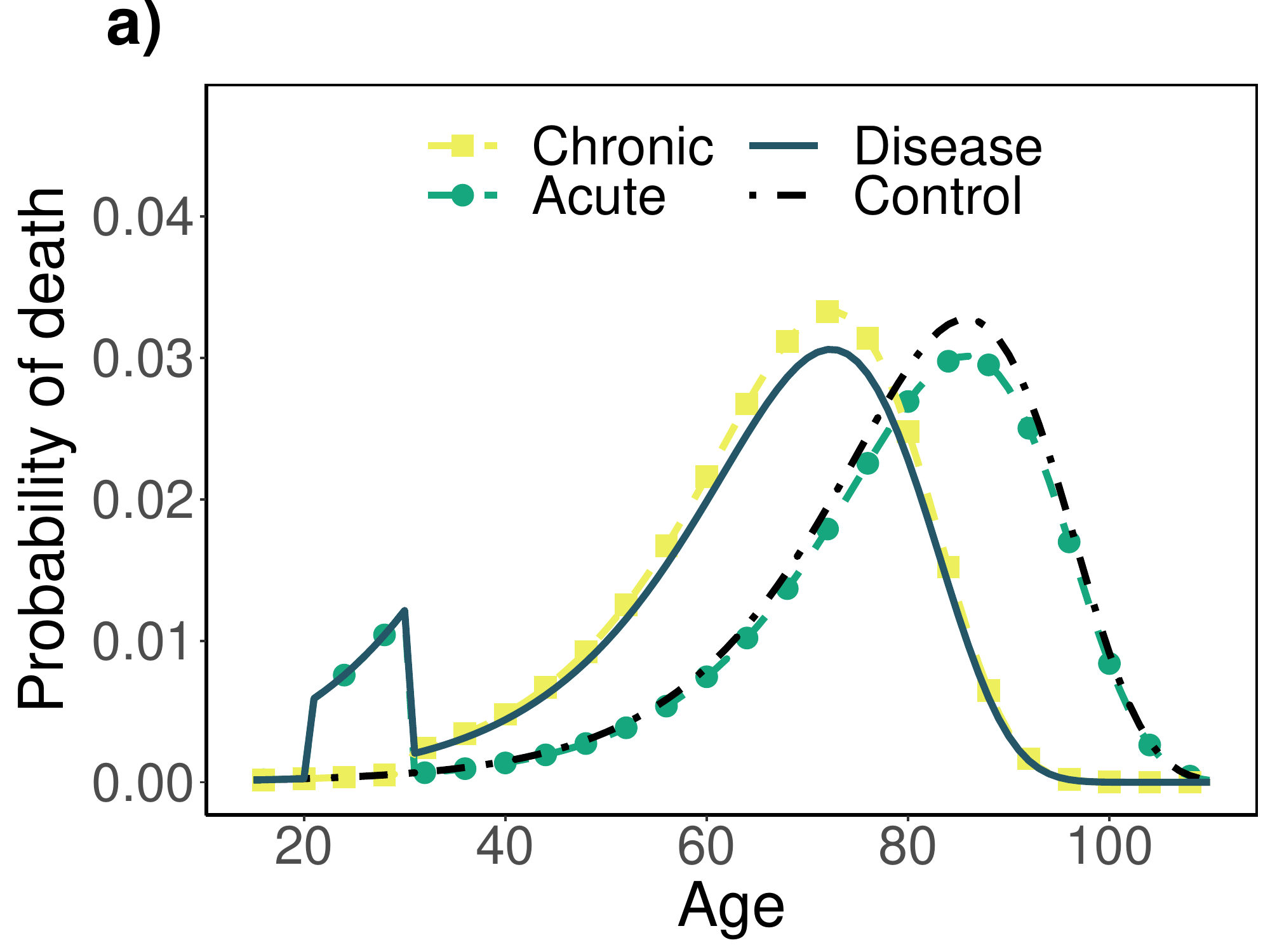}
    \includegraphics[width=.49\textwidth]{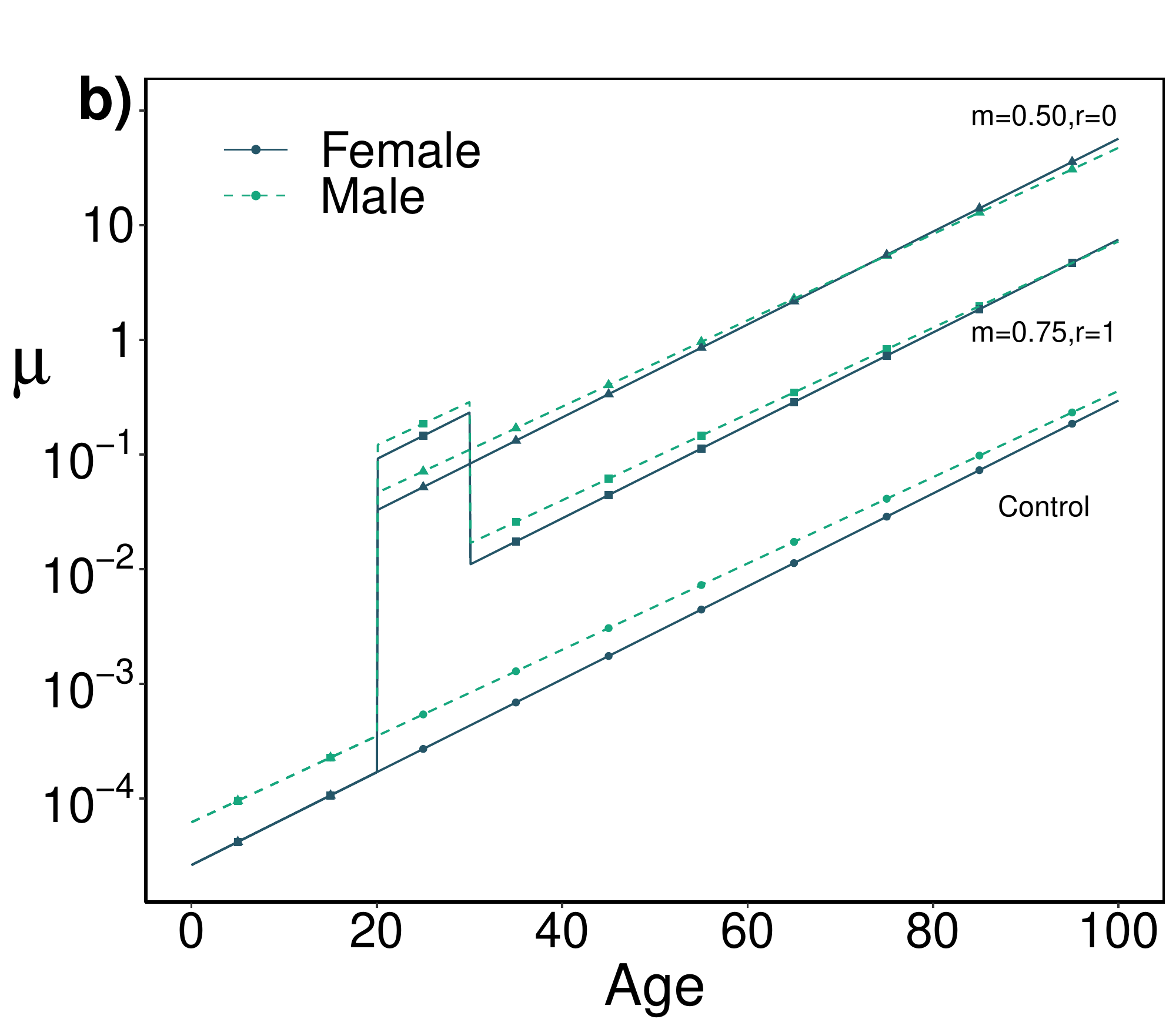} 
    \caption{ \textbf{Phenomenological model of disease.} \textbf{a) Acute and chronic phases.} Examples for the death age distributions. The disease has two effects (solid line): an acute effect causing a spike of deaths during the disease (ages 20-30) and a chronic effect after the disease that shifts the distribution towards younger ages (ages 30+). The acute phase is defined by having identical statistics to the disease until the end time, $t_{end}$. Conversely, the chronic phase is defined by having identical statistics to the disease after the end time, $t_{end}$. Deaths during the disease cause a screening effect, hence neither the chronic and disease nor the acute and control coincide after the disease. ($\tau=10$, $m=0.20$, $t_{on}=20$, $r=1$).
    \textbf{b) Sex effects of disease on mortality risk, $\mu$.} Males and females are known to have different survival rates, which are captured by our phenomenological model. Males (dashed green lines) have a higher baseline mortality (log-intercept) resulting in worse acute effects whereas females (solid blue lines) have higher slope such that they will eventually catch and exceed males for a sufficiently strong disease. From top to bottom: chronic disease, acute disease and control. The control fits of death probabilities from the Human Mortality Database, 2010 USA life tables \cite{hmd} are: $\mu_M = (6.0 \pm 0.3) \times 10^{-5}\exp{[(0.087\pm0.001)\times\text{Age}]}$ for males, and $\mu_F = (2.5 \pm 0.2) \times 10^{-5}\exp{[(0.094\pm0.001)\times\text{Age}]}$ for females.
    }
    \label{fig:pdfs}
\end{figure}

The probability density functions are illustrated in Fig.~\ref{fig:pdfs}a. The phenomenological model depends on survival parameters from Gompertz law which can be sex-specific -- this slightly changes the way males and females experience the disease, Fig.~\ref{fig:pdfs}b. Our model predicts that males will see higher short-term mortality rates during both the acute and chronic phases, due to their larger baseline mortality. If the disease is severe enough, females will eventually catch up and surpass the males in mortality rate due to their higher Gompertzian slope. Note that males and females also experience disease differently, with males tending to be more prone to infections \cite{Klein2016-vc}. This would correspond to different disease-specific parameters for males and females.

The loss of life for short and long effects are then computed using Eqn.~\ref{eq:dt} and substituting in for $p(t)$. Note that the loss of life has a non-trivial relationship between the full disease, $\langle \Delta t \rangle$ vs the short and long diseases,
\begin{align}
    \langle \Delta t \rangle &= \langle \Delta t \rangle_{short} + \langle \Delta t \rangle_{long}\frac{S_2(t_{end},m)}{S_2(t_{end},m=0)} \nonumber \\
    &= \langle \Delta t \rangle_{short} + \langle \Delta t \rangle_{long} - \langle \Delta t \rangle_{long}\bigg( 1 - \frac{S_2(t_{end},m)}{S_2(t_{end},m=0)}\bigg) \nonumber \\
    &\neq  \langle \Delta t \rangle_{short} + \langle \Delta t \rangle_{long} & \text{(overestimates effect)}
\end{align}
there is an additional corrective term that accounts for the fact that the chronic effects are under-estimated by the full disease due to deaths during the disease. That is, there would be more people to die if the acute portion of the disease never occurred (you can't die during both the acute phase and the chronic phase).  Note, however, that if the disease has low lethality then
\begin{align}
    \langle \Delta t \rangle &\approx \langle \Delta t \rangle_{short} + \langle \Delta t \rangle_{long} &\text{if}~S_2(t_{end},m)\approx S_2(t_{end},m=0).
\end{align}
Which is assured for sufficiently small $\tau$ or $m$.

\subsubsection{$p_{short}$ proof} \label{sec:shortproof} 
Here we derive Eqn.~\ref{eq:pshort}.

By definition, the short (acute) phase has the same statistics as the disease during the disease then reverts to control after the disease. The statistics are governed by the death age distribution hence,
\begin{align}
    p_{short}(t,m) = 
\begin{cases}
Z^{-1}_1p(t,m=0) & t < t_{on} \nonumber \\
Z^{-1}_2p(t,m) & t_{on} \leq t < t_{end} \nonumber \\
Z^{-1}_3p(t,m=0) & t \geq t_{end}.
\end{cases}
\end{align}
where we have included normalization constants for complete generality. 

By our assumption of identical statistics we should have $Z^{-1}_1p(t,m=0) = p(t < t_{on},m)$ and $Z^{-1}_1 = Z^{-1}_2 = 1$, but we also recover these here for completeness. Immediately we have $p(t < t_{on},m=0) = p(t < t_{on},m) = p_{short}(t < t_{on},m)$ as the hazard, $\mu$, is the same prior to the disease and the survival function is causal i.e.\ it doesn't see the disease coming in the future. Hence $Z_1 = 1$ and $p_{short}(t < t_{on},m) = p(t < t_{on},m)$.

Next we consider the survival at $t_{end}$. The short and complete diseases must have identical survival up until $t_{end}$,
\begin{align}
    S(t_{end},m) &= S_{short}(t_{end},m), \nonumber \\
    \int_{0}^{t_{end}} p(t,m) dt &= \int_{0}^{t_{end}} p_{short}(t,m) dt \nonumber \\
    &= Z^{-1}_2 \int_{0}^{t_{end}} p(t,m) dt \nonumber \\
\implies Z^{-1}_2 &= 1
\end{align}
where we have used the relation $S(t)=1-\int_0^t p(u) du$ (Eqn.~\ref{eq:sintp}).

Now we apply the normalization constraint. For $p_{short}$ to be a valid distribution we must have
\begin{align}
    \int_0^\infty p_{short}(t) &= 1 \nonumber \\
    &= \int_0^{t_{end}} p(t,m) dt + \int_{t_{end}}^{\infty}  Z^{-1}_3p(t,m=0) dt \nonumber \\
    &= (1-S(t_{end},m)) + Z^{-1}_3\bigg(\int_{0}^{\infty}  p(t,m=0) dt - \int_{0}^{t_{end}}  p(t,m=0) dt \bigg) \nonumber \\
    &= (1-S(t_{end},m)) + Z^{-1}_3\bigg(1 - (1 - S(t_{end},m=0) \bigg) \nonumber \\
\implies (1-S(t_{end},m)) + Z^{-1}_3S(t_{end},m=0) &= 1 \nonumber \\
\implies Z^{-1}_3 &= \frac{S(t_{end},m)}{S(t_{end},m=0)}.
\end{align}
Altogether we have
\begin{align}
    p_{short}(t,m) = 
\begin{cases}
p(t,m) & t < t_{on} \nonumber \\
p(t,m) & t_{on} \leq t < t_{end} \nonumber \\
p(t,m=0)\frac{S(t_{end},m)}{S(t_{end},m=0)} & t \geq t_{end}.
\end{cases}
\end{align}
\textbf{QED.}

\subsubsection{$p_{long}$ proof} \label{sec:longproof} 
Here we derive Eqn.~\ref{eq:plong}.

By definition, the long (chronic) phase has the same statistics as the control until the end of the disease state, at which point the statistics switch over to disease. The statistics are governed by the death age distribution hence,
\begin{align}
    p_{long}(t,m) = 
\begin{cases}
Z^{-1}_1p(t,m=0) & t < t_{end} \nonumber \\
Z^{-1}_2p(t,m) & t \geq t_{end}.
\end{cases}
\end{align}
where we have included normalization constants for complete generality.

Consider the survival at $t_{end}$. The long and control must have identical survival up until $t_{end}$,
\begin{align}
    S(t_{end},m=0) &= S_{long}(t_{end},m), \nonumber \\
    \int_{0}^{t_{end}} p(t,m=0) dt &= \int_{0}^{t_{end}} p_{long}(t,m) dt \nonumber \\
    &= Z^{-1}_1 \int_{0}^{t_{end}} p(t,m=0) dt \nonumber \\
\implies Z^{-1}_1 &= 1
\end{align}
where we have used the relation $S(t)=1-\int_0^t p(t) dt$ (Eqn.~\ref{eq:sintp}).

Now we apply the normalization constraint. For $p_{long}$ to be a valid distribution we must have
\begin{align}
    \int_0^\infty p_{long}(t) &= 1 \nonumber \\
    &= \int_0^{t_{end}} p(t,m=0) dt + \int_{t_{end}}^{\infty}  Z^{-1}_2p(t,m) dt \nonumber \\
    &= (1-S(t_{end},m=0)) + Z^{-1}_2\bigg(\int_{0}^{\infty}  p(t,m) dt - \int_{0}^{t_{end}}  p(t,m) dt \bigg) \nonumber \\
    &= (1-S(t_{end},m=0)) + Z^{-1}_2\bigg(1 - (1 - S(t_{end},m) \bigg) \nonumber \\
\implies (1-S(t_{end},m=0)) + Z^{-1}_2S(t_{end},m) &= 1 \nonumber \\
\implies Z^{-1}_2 &= \frac{S(t_{end},m=0)}{S(t_{end},m)}.
\end{align}
Altogether we have
\begin{align}
    p_{long}(t,m) = 
\begin{cases}
p(t,m=0) & t < t_{end} \nonumber \\
p(t,m)\frac{S(t_{end},m=0)}{S(t_{end},m)} & t \geq t_{end}.
\end{cases}
\end{align}
\textbf{QED.}

\subsubsection{$\Delta p_{death}$ Approximation}
In this section we obtain a result from the main text, specifically,
\begin{align}
    \Delta p_{death} 
    &\equiv \int_{t_{on}}^{t_{end}} \Delta p dt \approx \frac{m \tau \beta \mu_0}{\alpha f_{on}}, \label{eq:pdeath}
\end{align}
which is the difference between the probability of dying due to the acute phase minus the control. Note that $\mu_0$ is the control group hazard at the start of the disease.

We start by expanding $p$ for small $m$ and $\tau$. For $t_{on} \leq t < t_{end}$,
\begin{align}
    p(t,m) &= \mu(t,m)S(t,m) \nonumber \\
    &= b \bigg(1 + \frac{m}{f_{on}} \bigg)^{\beta/\alpha}e^{\beta t} S_1(t_{on})S_2(t) \nonumber \\
    &\approx be^{\beta t} \bigg(1 + \frac{\beta m}{\alpha f_{on}} \bigg)S_1(t_{on})(1 - \mu(t_{on},m)(t-t_{on})).
\end{align}
where $1 - \mu(t_{on},m)(t-t_{on})$ is the Taylor series of $S_2(t)$ for $t$ near $t_{on}$.

Now we compute the difference in probability of death during the disease,
\begin{align}
    \Delta p_{death} 
    &\equiv \int_{t_{on}}^{t_{end}} p(t,m)-p(t,m=0) dt \nonumber \\
    &\approx  \int_{t_{on}}^{t_{end}}  be^{\beta t} \bigg(1 + \frac{\beta m}{\alpha f_{on}} \bigg)S_1(t_{on})(1 - \mu(t_{on},m)(t-t_{on})) - be^{\beta t} S_1(t_{on})(1 - \mu(t_{on},m=0)(t-t_{on}))  dt \nonumber \\
    &=  \int_{t_{on}}^{t_{end}}  be^{\beta t} \frac{\beta m}{\alpha f_{on}} S_1(t_{on}) dt + \mathcal{O}(\tau^2) \nonumber \\
    &\approx S_1(t_{on}) b e^{\beta t_{on}} \frac{m}{\alpha f_{on}}(e^{\beta \tau}-1) \nonumber \\
    &\approx S_1(t_{on}) \frac{m \tau \beta}{\alpha f_{on}} \mu(t_{on},m=0).
\end{align}
If we condition on individuals living long enough to get the disease then $S_1(t_{on})=1$ and we have
\begin{align}
    \Delta p_{death} 
    &\approx \frac{m \tau \beta \mu_0}{\alpha f_{on}}.
\end{align}

\subsection{Acute effects}
Suppose that the disease group, $m>0$, experiences the disease but recovers completely after the end of the disease, meaning that they have identical survival statistics after the disease, Eqn.~\ref{eq:pshort}. The expected loss of lifespan can then be written as
\begin{align}
    \langle \Delta t \rangle_{short} &= \int_0^{\infty} t(p(t,m=0)-p_{short}(t,m)) dt \nonumber\\
    &= \int_0^{t_{on}} t(p(t,m=0)-p_{short}(t,m)) dt + \int_{t_{on}}^{t_{end}} t(p(t,m=0)-p_{short}(t,m)) dt  + \int_{t_{end}}^\infty t(p(t,m=0)-p_{short}(t,m)) dt  \nonumber \\
    &= \int_0^{t_{on}} t(p(t,m=0)-p(t,m)) dt+ \int_{t_{on}}^{t_{end}} t(p(t,m=0)-p(t,m)) dt + \int_{t_{end}}^\infty t(p(t,m=0)-p(t,m=0)\frac{S(t_{end},m)}{S(t_{end},m=0}) dt 
\end{align}
Using integration by parts, this can be re-written in terms of the survival function via Eqn.~\ref{eq:dtparts},
\begin{align}
    \langle \Delta t \rangle_{short} =& -t_{on}S(t_{on},m=0)+t_{on}S(t_{on},m)+\int_0^{t_{on}} S(t,m=0)-S(t,m) dt \nonumber \\
    &+ t_{on}S(t_{on},m=0) - t_{on}S(t_{on},m) - t_{end}S(t_{end},m=0) + t_{end}S(t_{end},m)+\int_{t_{on}}^{t_{end}} S(t,m=0)-S(t,m) dt  \nonumber \\
    &+ t_{end}S(t_{end},m=0) - t_{end}S(t_{end},m=0)\frac{S(t_{end},m)}{S(t_{end},m=0}) + \int_{t_{end}}^\infty S(t,m=0)-S(t,m=0)\frac{S(t_{end},m)}{S(t_{end},m=0)} dt \nonumber \\
    =& \int_0^{t_{on}} S(t,m=0)-S(t,m) dt +\int_{t_{on}}^{t_{end}} S(t,m=0)-S(t,m) dt  + \int_{t_{end}}^\infty S(t,m=0)-S(t,m=0)\frac{S(t_{end},m)}{S(t_{end},m=0)} dt 
\end{align}
It is convenient to work in $S_1$, $S_2$ and $S_3$ since it decouples the disease region from the rest of $S$ (Eqn.~\ref{eq:s1s2s3b}). Observe 
\begin{align}
    \langle \Delta t \rangle_{short} =& \int_{0}^{t_{on}} S_1(t,m=0)-S_1(t,m) dt + S_1\int_{t_{on}}^{t_{end}} S_2(t,m=0)-S_2(t,m) dt + \nonumber \\
    &+S_1\int_{t_{end}}^{\infty} S_2(t_{end},m=0)S_3(t,m=0)-S_2(t_{end},m)S_3(t,m=0) dt.
\end{align}
The survival components are identical before, $S_1(m=0)=S_1(m)$, and after, $S_3(m=0)=S_3(m)$, the disease which leads to
\begin{align}
    \langle \Delta t \rangle_{short} &= S_1(t_{on})\int_{t_{on}}^{t_{end}} S_2(t,m=0)-S_2(t,m) dt + S_1(t_{on})\big(S_2(t_{end},m=0)-S_2(t_{end},m)\big)\int_{t_{end}}^{\infty} S_3(t,m=0) dt.    
\end{align}
These integrals include double exponential functions which are non-trivial to compute analytically. In the main text we present numerical integration results in Fig.~4b. A perturbative approach reveals useful insights. Suppose the disease occurs for a short period, $\tau$, then we can approximate
\begin{align}
    S_2(t,m) &\approx S_2(t_{end},m) + \frac{dS_2}{dt}\bigg|_{t_{end}}(t-t_{end}) = S_2(t_{end},m) - \mu(t_{end},m)S_2(t_{end},m)(t-t_{end}) \label{eq:ds2dtau}
\end{align}
we then have 
\begin{align}
    \langle \Delta t \rangle_{short} &\approx S_1(t_{on})(S_2(t_{end},m=0)-S_2(t_{end},m))\tau +  S_1(\mu(t_{end},m)S_2(t_{end},m)-\mu(t_{end},m=0)S_2(t_{end},m=0))(-\tau^2/2) \nonumber \\
    &+ S_1(t_{on})\big(S_2(t_{end},m=0)-S_2(t_{end},m)\big)\int_{t_{on}}^{\infty} S_3(t,m=0) dt \nonumber \\
    &\approx S_1(t_{on})(S_2(t_{end},m=0)-S_2(t_{end},m))\bigg(\tau +  \int_{t_{end}}^{\infty} S_3(t,m=0) dt \bigg)
\end{align}
where we've dropped $O(\tau^2)$ terms due to our assumption that $\tau$ is small. Note that $S_3(t_{end})=1$ and $S_3(\infty)=0$. The integral can be approximated as a step function centered at the characteristic time, Eqn.~\ref{eq:sint}, to yield
\begin{align}
   \Aboxed{ \langle \Delta t \rangle_{short} &\approx S_1(t_{on}) \bigg(S_2(t_{end},m=0)-S_2(t_{end},m) \bigg)\bigg(\tau + (t_e - t_{end}) \bigg).} \label{eq:tshort}
\end{align}
where $t_{e}$ is given by Eqn.~\ref{eq:te} with $t_r\equiv t_{end}$. Note that $S_1 = 1$ conditions on the population having lived long enough to get the disease, which is standard practice (e.g.\ excess lifetime risk \cite{Thomas1992-lt}).

\subsubsection{Weak disease, small $m$ and $\tau$}
Starting from Eqn.~\ref{eq:tshort}, which assumed small $\tau$, if we further assume $m$ is small then we can write
\begin{align}
    S_2(t,m) &\approx S_2(t,m=0) + m\frac{\partial S_2}{\partial m}\bigg|_{m=0}  \nonumber \\
    &= S_2(t,m=0) - \frac{m\mu(t,m=0)S_2(t,m=0)}{\alpha f_{on}}(1-e^{-\beta(t-t_{on})})
\end{align}
where $f_{on}\equiv ae^{\beta t_{on}}$. Thus we can write
\begin{align}
    \langle \Delta t \rangle_{short} &\approx S_1(t_{on}) \bigg( S_2(t_{end},m=0)-S_2(t_{end},m=0) + \frac{m\mu(t_{end},m=0)S_2(t_{end},m=0)}{\alpha f_{on}}(1-e^{-\beta(t_{end}-t_{on})}) \bigg) \bigg(\tau +  (t_{e}-t_{end}) \bigg).
\end{align}
Next we apply our earlier condition of small $\tau$ to expand the exponential,
\begin{align}
    \langle \Delta t \rangle_{short} &\approx S_1(t_{on})S_2(t_{end},m=0)\frac{m\beta\tau\mu(t_{end},m=0)}{\alpha f_{on}}\bigg(\tau + (t_{e}-t_{end})  \bigg). \label{eq:tshortapprox}
\end{align}
Substituting in Eqn.~\ref{eq:te} we get
\begin{align}
   \langle \Delta t \rangle_{short} &\approx S_1(t_{on})S_2(t_{end},m=0)\frac{m\beta\tau\mu(t_{end},m=0)}{\alpha f_{on}}\bigg(\tau + \bigg(-t_{end} + \frac{1}{\beta}\ln{\bigg(\frac{\beta}{b}+e^{\beta t_{end}}\bigg)} \bigg)  \bigg) \nonumber \\
   &= S_1(t_{on})S_2(t_{end},m=0)\frac{m\beta\tau\mu(t_{end},m=0)}{\alpha f_{on}}\bigg(\tau + \frac{1}{\beta}\ln{\bigg(\frac{\beta}{\mu_{end}}+1\bigg)}  \bigg).
\end{align}
We have already assumed small $\tau$. To lowest order in $\tau$ this is
\begin{align}
   \Aboxed{ \langle \Delta t \rangle_{short} &\approx S_1(t_{on})S_2(t_{end},m=0)\frac{m\tau}{f_{on}}\frac{\mu(t_{end},m=0)}{\alpha}\ln{\bigg(1+\frac{\beta}{\mu(t_{end},m=0)}\bigg)}. } \label{eq:dtshortapprox}
\end{align}
where $S_2(t_{end},m=0) = 1 - \mu(t_{on},m=0)\tau \approx 1$ can also be neglected in the strict limit -- though we leave it in place (see below). Furthermore, note that $S_1(t_{on})=1$ if we condition on individuals being alive at the start of the disease. 

\subsection{Chronic effects} \label{sec:chronic}
Suppose that the disease group, $m>0$, has identical survival to the control population until time $t_{end}$ when they begin to follow the chronic survival statistics of the disease, Eqn.~\ref{eq:plong}. The expected loss of lifespan can then be written as
\begin{align}
    \langle \Delta t \rangle_{long} &= \int_0^{t_{end}} t(p(t,m=0)-p_{long}(t,m)) dt + \int_{t_{end}}^\infty t(p(t,m=0)-p(t,m)) dt \nonumber \\
    &= 0 + \int_{t_{end}}^\infty t(p(t,m=0)-p_{long}(t,m=1)) dt 
\end{align}
By definition (Eqn~\ref{eq:plong}),
\begin{align}
    \langle \Delta t \rangle_{long} &= \int_{t_{end}}^\infty t(p(t,m=0)-p(t,m)\frac{S(t_{end},m=0)}{S(t_{end},m)}) dt  
\end{align}
The ratio of survival functions ensures that $p(t,m)$ is properly normalized.
Using Eqn.~\ref{eq:s1s2s3b} we can write
\begin{align}
     \langle \Delta t \rangle_{long} &= S_1(t_{on},m=0)S_2(t_{end},m=0)\int_{t_{end}}^\infty S_3(t,m=0)-S_3(t,m) dt.  \label{eq:dtlongdef}
\end{align}
Eqn.~\ref{eq:dtlongdef} can be solved numerically. We seek a more insightful expression. The residual effects of the disease are often weak, motivating us to expand in the effective gain in age, $\delta$ (Section~\ref{sec:ba}).

We expand $S_3$ in small gained age, $\delta$, using Eqn.~\ref{eq:ba},
\begin{align}
    S_3(m) &= \exp{\bigg( -\frac{b}{\beta}\big(1+\frac{\Delta f}{{f_{end}}}\big)^{\beta/\alpha}\big(e^{\beta t} - e^{\beta t_{end}}\big) \bigg)} \nonumber \\
    &= \exp{\bigg( -\frac{b}{\beta}e^{\beta\delta}\big(e^{\beta t} - e^{\beta t_{end}}\big) \bigg)} \nonumber \\
    &\approx S_3(m=0) + \delta \frac{\partial S_3}{\partial \delta}\bigg|_{\delta=0} \nonumber \\
    S_3(m) &\approx S_3(m=0) + \delta\bigg( \frac{d}{dt}S_3(m=0) + be^{\beta t_{end}}S_3(m=0) \bigg)
\end{align}
where we have Taylor expanded near $\delta =0$. Note that $m=0$ if and only if $\delta = 0$. Eqn.~\ref{eq:dtlongdef} then simplifies,
\begin{align}
    \langle \Delta t \rangle_{long} &\approx S_1(t_{on},m=0)S_2(t_{end},m=0)\int_{t_{end}}^\infty S_3(t,m=0)- S_3(t,m=0) - \delta\bigg( \frac{d}{dt}S_3(t,m=0) + be^{\beta t_{end}}S_3(t,m=0) \bigg) dt \nonumber \\
    &= \delta S_1(t_{on},m=0)S_2(t_{end},m=0)\bigg( -S_3(t, m=0)\bigg|_{ts}^\infty - \mu(t_{end},m=0) \int_{t_{end}}^\infty S_3(t,m=0)dt \bigg)\nonumber \\
    &= \delta S_1(t_{on},m=0)S_2(t_{end},m=0)\bigg( 1 - \mu(t_{end},m=0) \int_{t_{end}}^\infty S_3(t,m=0)dt \bigg).
\end{align}
We characterize the integral, $\int_{t_{end}}^\infty S_3(t,m=0)dt$ in Section~\ref{sec:ints3}, resulting in a pair of constraints,
\begin{align}
    0 < & \mu(t_{end},m=0)\int_{t_{end}}^\infty S_3(t,m=0)dt < 1,~\text{and} \nonumber \\
    0 < & \mu(t_{end},m=0)\int_{t_{end}}^\infty S_3(t,m=0)dt < \frac{1}{2}\exp{\bigg(\frac{\mu(t_{end},m=0)}{2\beta}\bigg)}\sqrt{\frac{2\pi\mu(t_{end},m=0)}{\beta}}.
\end{align}
At young ages, the upper limit is $\ll 1$, for example at age $t_{end}=20$ it is 0.07.

In general, $\mu(t_{end},m=0)\int_{t_{end}}^\infty S_3(t,m=0)dt$ does not contribute much at younger ages but becomes increasingly closer to $1$ at older ages. A simple way to account for this behaviour is to approximate the integral by a step function (Section~\ref{sec:timescales}). This assumes the survival function drops very rapidly, which is true for old ages (for young ages the term is small and can be ignored). With this approximation we have,
\begin{align}
    \langle \Delta t \rangle_{long} &\approx \delta S_1(t_{on},m=0)S_2(t_{end},m=0)\bigg( 1 - \mu(t_{end},m=0) \int_{t_{end}}^\infty I(t_{end} < t_{e})  dt \bigg) \nonumber\\
    \langle \Delta t \rangle_{long} &\approx \delta S_1(t_{on},m=0)S_2(t_{end},m=0)\bigg( 1 - \mu(t_{end},m=0) (t_{e} - t_{end}) \bigg)
\end{align}
where $I$ is an indicator function ($1$: true, $0$: false), and $t_{e}$ is given by Eqn.~\ref{eq:te}. Typically we will assume the individual lived long enough to get the disease, $S_1=1$. In terms of model parameters we can substitute for $\delta$ (Eqn.~\ref{eq:delta}) to get
\begin{align}
    \Aboxed{ \langle \Delta t \rangle_{long} &\approx S_1(t_{on},m=0)S_2(t_{end},m=0)\frac{1}{\alpha}\ln{\bigg( 1 + \frac{\Delta f}{f_{end}} \bigg)}\bigg( 1 - \mu(t_{end},m=0) (t_{e} - t_{end}) \bigg) }. \label{eq:dtlong}
\end{align}
Recall $f_{end}$ is the control FI at time $t_{end} = t_{on}+\tau$. The approximation is exact for weak diseases (equivalently: small $\delta$, small $m$, or small $\tau$ with $r=1$). Note that the second term can never be greater than $1$ and is nearly $0$ for young $t_{end}$.

\subsubsection{Characterizing $\int_{t_{end}}^\infty S_3(t,m=0)dt$} \label{sec:ints3} 
The integral, $\int_{t_{end}}^\infty S_3(t,m=0)dt$, can be characterized by upper and lower bounds. For lower bound, $S_3(t) > 0$ hence the integral must positive. For upper bound, we can expand the inner exponential as a polynomial. Using the definition of $S_3$, Eqn.~\ref{eq:s1s2s3}, and Gompertz law for the hazard we have,
\begin{align}
    \int_{t_{end}}^\infty \exp{\bigg(\frac{-b}{\beta}(e^{\beta t}-e^{\beta t_{end}})\bigg)} dt &\lesssim \int_{t_{end}}^\infty \exp{\bigg(\frac{-be^{\beta t_{end}}}{\beta}(1+\beta(t-t_{end})+\frac{\beta^2}{2}(t-t_{end})^2-1)\bigg)} dt 
\end{align}
where we use $\lesssim$ to indicate that the right hand side is always larger and will be approximately equal for large $t_{end}$. Large $t_{end}$ ensures the integrand drops rapidly and hence $t\approx t_{end}$ dominates the integral. To linear order we have,
\begin{align}
    \int_{t_{end}}^\infty \exp{(-b e^{\beta t_{end}}(t-t_{end})))} dt &= \exp{(b e^{\beta t_{end}}t_{end})} \frac{\exp{(-b e^{\beta t_{end}}t)}}{-be^{\beta t_{end}}}\bigg|_{t_{end}}^\infty \nonumber \\
    &= \frac{1}{\mu(t_{end},m=0)},
\end{align}
noting $\mu(t_{end},m=0) = b e^{\beta t_{end}}$ from Gompertz' law.
This constrains
\begin{align}
    0 < & \mu(t_{end},m=0)\int_{t_{end}}^\infty S_3(t,m=0)dt < 1.
\end{align}
If we also consider the quadratic term we can set a different upper limit,
\begin{align}
    \int_{t_{end}}^\infty \exp{(-b e^{\beta t_{end}}(t-t_{end} + \frac{\beta}{2}(t-t_{end})^2)))} dt &= \int_{0}^\infty \exp{(-\frac{\beta}{2} b e^{\beta t_{end}}( \frac{2}{\beta} u + u^2)))} du \nonumber \\
    &= \exp{\bigg(\frac{\mu(t_{end},m=0)}{2\beta}\bigg)} \int_{0}^\infty \exp{(-\frac{1}{2}\mu(t_{end},m=0)\beta (u+1/\beta)^2 )} du \nonumber \\
    &= \exp{\bigg(\frac{\mu(t_{end},m=0)}{2\beta}\bigg)}\sqrt{\frac{2\pi}{\mu(t_{end},m=0)\beta}} \bigg(1-\Theta(0; -\frac{1}{\beta}, \frac{1}{\sqrt{\mu(t_{end},m=0)\beta}}) \bigg)
\end{align}
where $\Theta$ is the integral of a normal distribution with mean $-1/\beta$ and standard deviation $(\mu(t_{end},m=0)\beta)^{-1/2}$. $\Theta \to 1$ as $\mu(t_{end},m=0)\to\infty$ hence the integral becomes small at older $t_{end}$. Regardless, we know that the normal distribution is symmetrical about the mean and $\beta > 0$ hence $\Theta(0) > 0.5$ and we must have,
\begin{align}
    0 < & \mu(t_{end},m=0)\int_{t_{end}}^\infty S_3(t,m=0)dt < \frac{1}{2}\exp{\bigg(\frac{\mu(t_{end},m=0)}{2\beta}\bigg)}\sqrt{\frac{2\pi\mu(t_{end},m=0)}{\beta}}
\end{align}
as an additional constraint. At young ages, the upper limit is $< 1$, for example at age $t_{end}=20$ it is 0.07.

\subsubsection{Weak disease, small $m$ and small $\tau$}
If the disease is sufficiently weak we can make a further approximation of Eqn.~\ref{eq:dtlong}. Implicitly, by assuming small $\delta$ we've already assumed either small $m$ or simultaneously small $\tau$ and $r=1$. Observe we have,
\begin{align}
    \langle \Delta t \rangle_{long} &\approx S_1(t_{on},m=0)S_2(t_{end},m=0)\frac{1}{\alpha}\ln{\bigg( 1 + \frac{m(e^{\alpha\tau}-r)}{f_{end}} \bigg)}\bigg( 1 - \mu(t_{end},m=0) (t_{e} - t_{end}) \bigg). 
\end{align}
Substituting in Eqn.~\ref{eq:te} for $t_e$,
\begin{align}
    \langle \Delta t \rangle_{long} &\approx S_1(t_{on},m=0)S_2(t_{end},m=0)\frac{1}{\alpha}\ln{\bigg( 1 + \frac{m(e^{\alpha\tau}-r)}{f_{end}} \bigg)}\bigg( 1 - \frac{\mu(t_{end},m=0)}{\beta} \ln{\bigg( 1 + \frac{\beta}{\mu(t_{end},m=0)} \bigg)} \bigg). 
\end{align}
For small $m$ we can then expand the logarithm,
\begin{align}
    \langle \Delta t \rangle_{long} &\approx S_1(t_{on},m=0)S_2(t_{end},m=0)\frac{1}{\alpha}\bigg( \frac{m(e^{\alpha\tau}-r)e^{-\alpha\tau}}{a e^{\alpha t_{on}}} \bigg)\bigg( 1 - \frac{\mu(t_{end},m=0)}{\beta} \ln{\bigg( 1 + \frac{\beta}{\mu(t_{end},m=0)} \bigg)} \bigg).
\end{align}
If we further assume small $\tau$ we see,
\begin{align}
    \Aboxed{ \langle \Delta t \rangle_{long} &\approx  S_1(t_{on},m=0)S_2(t_{end},m=0)\bigg( \frac{m \tau}{f_{on}} \big(r + \frac{1-r}{\alpha\tau} \big) \bigg)\bigg( 1 - \frac{\mu(t_{end},m=0)}{\beta} \ln{\bigg( 1 + \frac{\beta}{\mu(t_{end},m=0)} \bigg)} \bigg). } \label{eq:dtlongapprox}
\end{align}
If we consider only individuals whom survived to get the disease then $S_1(t_{on})=1$. We can also approximate $S_2(t_{end}) \approx 1 - \mu(t_{on})\tau \approx 1$, although we leave it in anticipation of Section~\ref{sec:ratio}. The $\mu(t_{end},m=0)$ term is small for young $t_{on}$ and can be neglected but reaches unity for the very old, as discussed above (recall that we used $\int_{t_{end}}^\infty S_3(t) dt \approx t_{e} - t_{end}$).

\subsection{Acute-Chronic ratio} \label{sec:ratio}
We consider the ratio of acute to chronic disease effects,
\begin{align}
    \frac{\langle \Delta t \rangle_{short}}{\langle \Delta t \rangle_{long}} &= \frac{S_1(t_{on},m=0)\int_{t_{on}}^{t_{end}} S_2(t,m=0)-S_2(t,m) dt + S_1(t_{on},m=0)\big(S_2(t_{end},m=0)-S_2(t_{end},m)\big)\int_{t_{end}}^{\infty} S_3(t,m=0) dt}{S_1(t_{on},m=0)S_2(t_{end},m=0)\int_{t_{end}}^\infty S_3(t,m=0)-S_3(t,m) dt} \nonumber \\
    &=\frac{\int_{t_{on}}^{t_{end}} S_2(t,m=0)-S_2(t,m) dt}{S_2(t_{end},m=0)\int_{t_{end}}^\infty S_3(t,m=0)-S_3(t,m) dt} + \frac{\big(S_2(t_{end},m=0)-S_2(t_{end},m)\big)\int_{t_{end}}^{\infty} S_3(t,m=0) dt}{S_2(t_{end},m=0)\int_{t_{end}}^\infty S_3(t,m=0)-S_3(t,m) dt}
\end{align}

For small $\tau$ and $m$ we can use Eqn.~\ref{eq:dtshortapprox} and Eqn.~\ref{eq:dtlongapprox},
\begin{align}
    \frac{\langle \Delta t \rangle_{short}}{\langle \Delta t \rangle_{long}} &\approx \frac{ S_1(t_{on})S_2(t_{end},m=0)\frac{m\tau}{f_{on}}\frac{\mu(t_{end},m=0)}{\alpha}\ln{\bigg(1+\frac{\beta}{\mu(t_{end},m=0)}\bigg)} }{ S_1(t_{on})S_2(t_{end},m=0)\bigg( \frac{m \tau}{f_{on}} \big(r + \frac{1-r}{\alpha\tau} \big) \bigg)\bigg( 1 - \frac{\mu(t_{end},m=0)}{\beta} \ln{\bigg( 1 + \frac{\beta}{\mu(t_{end},m=0)} \bigg)} } \nonumber \\
\Aboxed{ \frac{\langle \Delta t \rangle_{short}}{\langle \Delta t \rangle_{long}} &\approx \frac{ \beta\mu(t_{end},m=0)\ln{\bigg(1+\frac{\beta}{\mu(t_{end},m=0)}\bigg)}  }{\alpha \bigg(r + \frac{1-r}{\alpha\tau} \bigg)\bigg( \beta - \mu(t_{end},m=0) \ln{\bigg( 1 + \frac{\beta}{\mu(t_{end},m=0)} \bigg)} \bigg)}. } \label{eq:ratio}
\end{align}
As discussed in Section~\ref{sec:chronic}, the $\mu$ term in the denominator is always positive and $< 1$, increasing from a small correction in young ages to order unity by approximately age $100$. Hence dropping this term will give a lower limit that's tight (good) at younger ages but poor at older ages.
\bibliography{ref}  